\documentclass[a4paper]{PoS}
\usepackage{multirow,float}	
\usepackage{array,booktabs,arydshln,xcolor}
\newcommand\VRule[1][\arrayrulewidth]{\vrule width #1}

\title{Time resolution analysis of detectors based on plastic scintillators coupled to silicon photomultipliers}

\ShortTitle{Time resolution analysis of trigger counters}

\author{\speaker{Marco Alberto Ayala Torres}, Luis Manuel Monta\~{n}o Zetina and Marcos Fontaine S\'anchez \\
        Center for Research and Advanced Studies~(CINVESTAV)\\
        Physics Department, Mexico\\
        E-mail: \email{amayala@fis.cinvestav.mx}, \email{lmontano@fis.cinvestav.mx}, \email{mfontaine@fis.cinvestav.mx}}

\abstract{
The performance of several trigger counters based on plastic scintillators with 
silicon photomultiplier readout is investigated with cosmic rays. Efficiency 
and time resolution are measured using digital waveform analysis. The obtained 
results are relevant for trigger subsystems of \textbf{M}ulti-\textbf{P}urpose \textbf{D}etector~(MPD) 
and \textbf{B}aryonic \textbf{M}atter \textbf{at} the \textbf{N}uclotron~(BM@N) at the NICA heavy-ion collider. 
The results show very high efficiency and good timing performance of the counters.
}

\FullConference{7th Annual Conference on Large Hadron Collider Physics - LHCP2019\\
		20-25 May, 2019\\
		Puebla, Mexico \\
		PoS (LHCP2019) 062\\
		Published on: 2019 December 04\\
		DOI: https://doi.org/10.22323/1.350.0062
		}

\begin{document}

\section{Introduction} 
In modern high energy physic experiments, Silicon Photomultipliers~(SiPMs) have the potential to replace traditional Photomultiplier Tubes~(PMTs) in many applications, such as trigger detectors, time-of-flight hodoscopes, and calorimetry. Compared to PMTs, 
the SiPMs sensors provide similarly high photon detection efficiency and very good time resolution,  but besides, they offer compactness, ability to operate in a magnetic field and low power consumption. On the other hand, a single SiPM has a relatively small sensitive area, and therefore arrays of many SiPMs are required in applications where scintillators have large areas. 
Because the event rate in high energy physics experiments is very high, trigger detectors are typically based on either fast plastic scintillators or Cherenkov radiators coupled to PMTs or SiPMs. Basic requirements for such trigger detectors are very good efficiency of particle detection and excellent time resolution.

\begin{figure}[b]
\begin{center}
\includegraphics[width=0.85\textwidth]{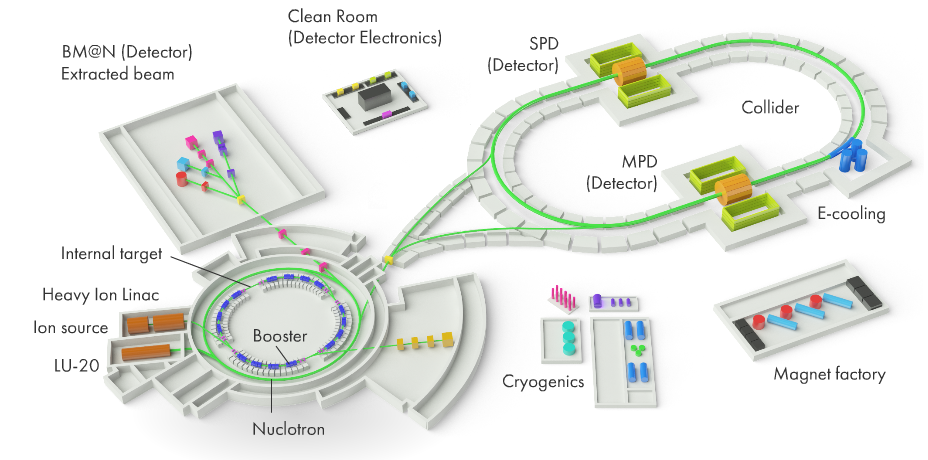}\\
\caption{The principal components of the NICA accelerator complex~(taken from http://nica.jinr.ru/complex.php).}
\label{nica}
\end{center}
\end{figure}

In this study, we evaluated the response of several simple scintillation counters to cosmic ray particles. The choice of scintillators and SiPMs was based on their availability and relevance to the trigger detector systems in the experiments  \textbf{M}ulti-\textbf{P}urpose \textbf{D}etector~(MPD) and \textbf{B}aryonic \textbf{M}atter at the \textbf{N}uclotron~(BM@N) at \textbf{N}uclotron based \textbf{I}on \textbf{C}ollider f\textbf{A}cility~(NICA). NICA is a modern accelerator complex at Joint Institute for Nuclear Research~(JINR) in Dubna, Russia~(see fig.~\ref{nica} for details). The start of the collider operation is expected in 2021.

Three detectors will cover the NICA project experimental program: BM@N, MPD, and the Spin Physics Detector~(SPD)~\cite{3stages}. At the first stage, the Nuclotron complex will be upgraded to accelerate high-intensity heavy ions beams to the energies~$1-6~GeV/nucleon$. These beams will be transported to the fixed target experiment BM@N. At the second stage, nucleus-nucleus collisions will be studied by the collider experiment MPD. At the last stage, beams of polarized protons and deuterons will allow spin physics studies by the SPD experiment.

In the collider mode the center-of-mass~(c.m.s.) energy will reach $\sqrt{s}_{NN} = 4~GeV-11~GeV$ for $Au+Au$ and up to $27~GeV$ for $p+p$ collisions. Thus, the experimental program of NICA covers two areas: the study of the hot strongly interacting nuclear matter at highest baryonic densities and also research in proton and deuteron spin physics~\cite{rfqNICA}.

\subsection{Timing trigger detectors in MPD}

The efficiency of the interaction trigger in MPD provided by the \textbf{F}ast \textbf{F}orward \textbf{D}etector sub-system~(FFD) will be close to 100\% for central collisions of heavy nuclei, such as $Au+Au$, but significantly lower for peripheral collisions of heavy nuclei, proton-proton collisions, and collisions of light nuclei~\cite{FFD}. This reduction of efficiency is due to limited acceptance of the FFD and much lower particle multiplicity in collisions of protons or light nuclei.   

\begin{figure}[htb]
\begin{center}
\includegraphics[width=0.4\textwidth]{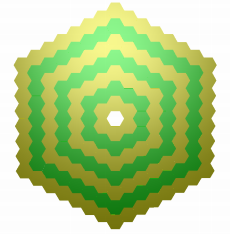}
\caption{The geometry of the Beam-Beam Counter Detector~(BE-BE, from MEXnICA collaboration).}
\label{BEBEhex}
\end{center}
\end{figure}

To increase the efficiency of low multiplicity events, the MEXnICA Group of the MPD collaboration proposed to include the \textbf{Be}am-\textbf{Be}am Counter Detector~(BE-BE), with similar performance to the FFD, but with larger acceptance~\cite{1Mario}. 
The proposed BE-BE detector is based on an array of hexagonal plastic scintillators (fig.~\ref{BEBEhex}) and light
sensors. Using SiPMs as the light sensors would be the best option.
The BE-BE detector will provide not only the Level-0 trigger but can enhance MPD capability to determine the centrality of the events and improve reaction plane resolution. Also, similar to the FFD, the BE-BE detector will provide the start signal for the time of flight measurements.

\subsection{The BM@N Barrel Detector~(BD)}

BM@N is the first experiment at the accelerator complex of NICA~\cite{BMAN}. The experiment is designed to study the interactions of relativistic heavy-ion beams with fixed targets. The Nuclotron will provide a variety of beams from protons to gold ions with energies ranging from $1$ to $6~GeV$ per nucleon. The BM@N spectrometer consists of several tracking, time-of-flight, and calorimeter sub-systems which will measure charged particle spectra, hyperon, and hypernuclei production and anisotropic flow in nuclear collisions.

\begin{figure}[t]
\begin{center}
\includegraphics[width=0.322\textwidth]{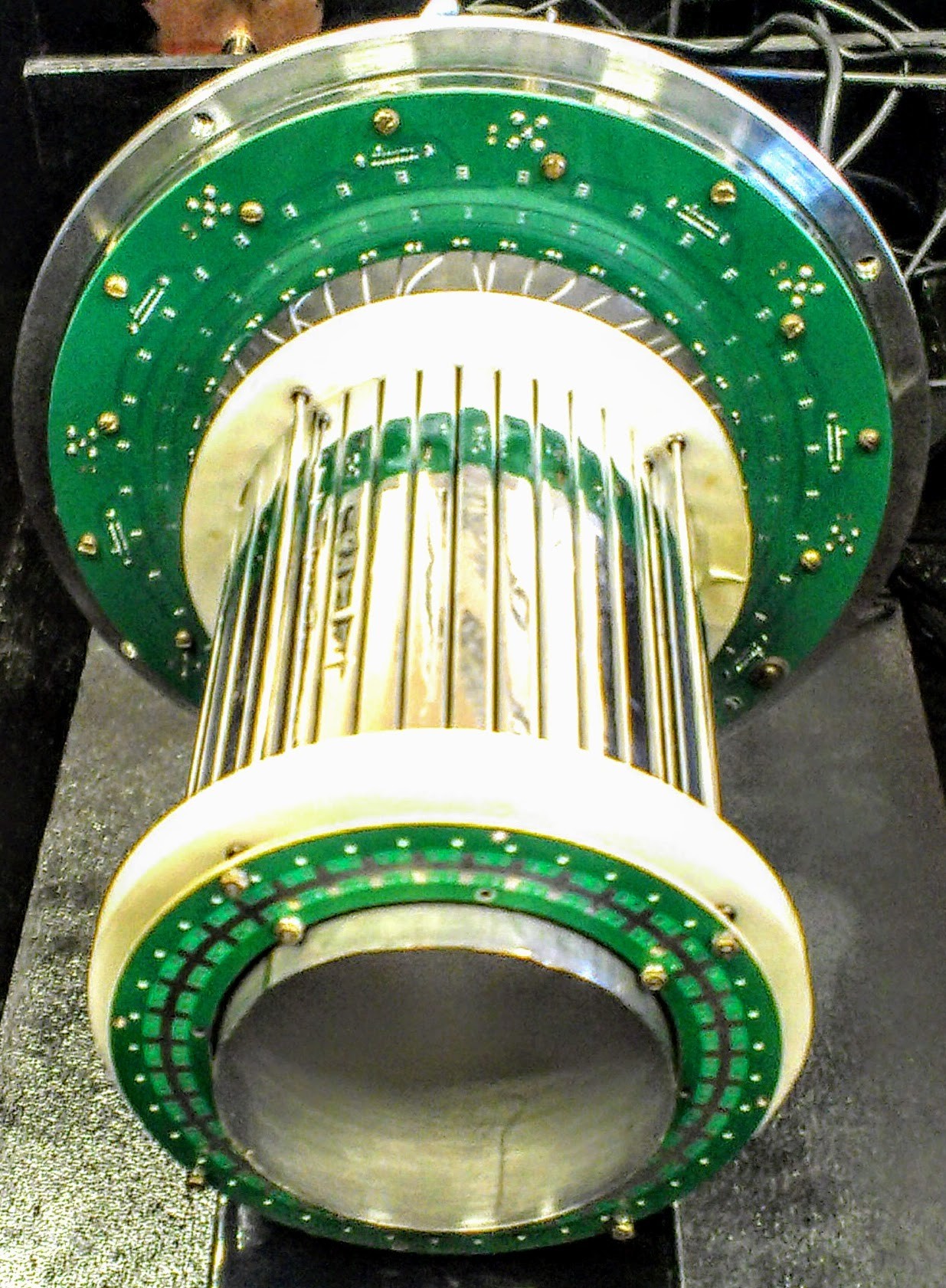}\hspace{5mm}
\includegraphics[width=0.56\textwidth]{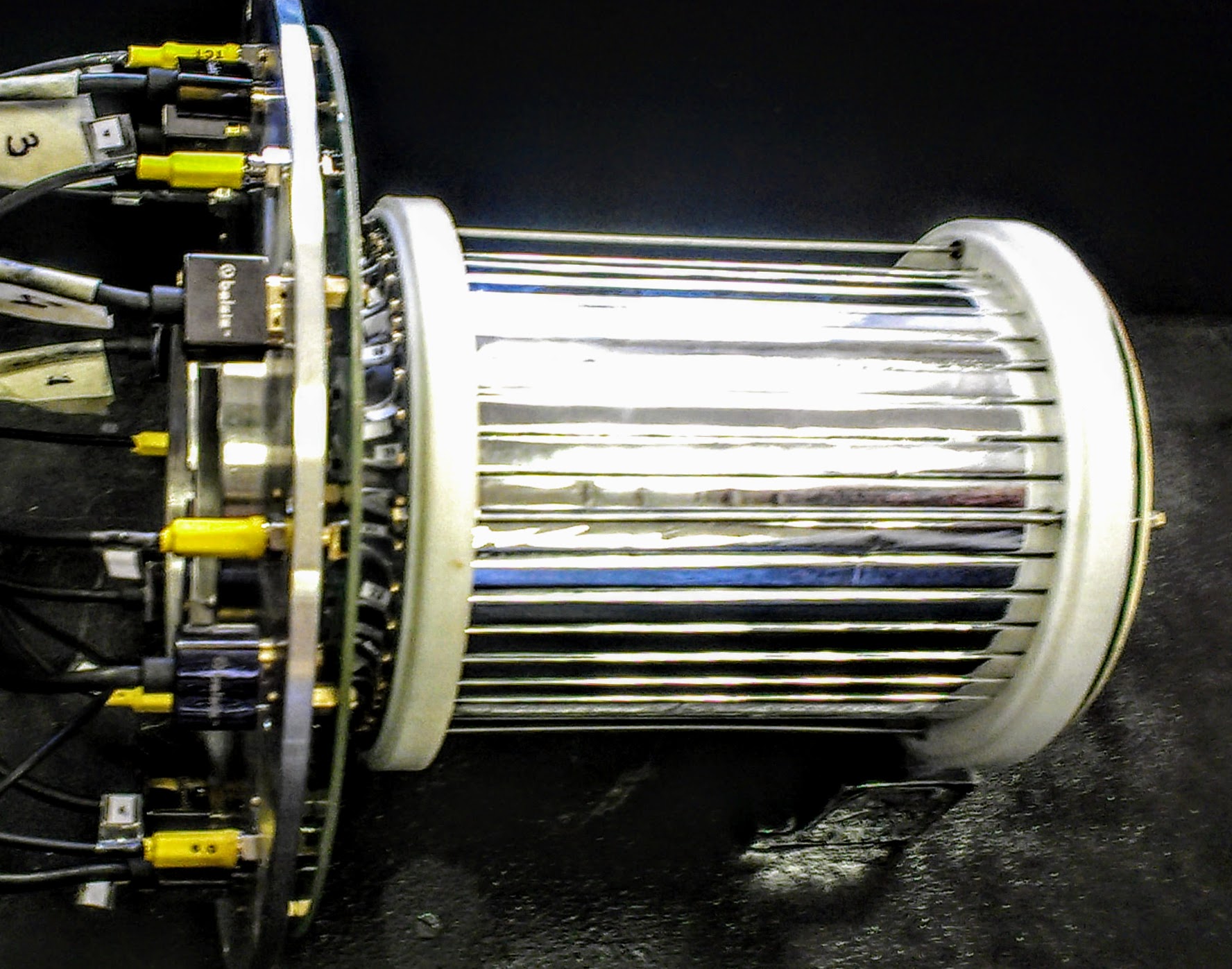} \\
\caption{The BM@N Barrel Detector~(BD).}
\label{BDpics}
\end{center}
\end{figure}

The BD is located in the target area of the spectrometer. It surrounds the target and consists of 40 strips $150\times7\times7~mm^3$ made of BC418 scintillator and wrapped in Al-mylar~(see fig.~\ref{BDpics}). Light from each strip is detected on one end by a single SiPM Sensl Micro FC-60035-SMT. Active area $6\times6~mm^2$ of the SiPM matches $7\times7~mm^2$ cross-section of the strip providing excellent light collection efficiency.

In the BM@N experiment, the BD should register charged particles emitted in the nuclei-nuclei collisions at large angles close to the target. The number of detected hits on average depends on the impact parameter of the collision; it is used to select the centrality of the collisions. Therefore, high efficiency for the minimum ionizing particles is the most important requirement for the BD detector.  
When the BD was finally mounted, there were some scintillator strips and SiPMs were leftover and used for this study. 
In addition to the configuration, in which light is collected by single SiPMs from one end of the strips, we made time resolution measurements for the configuration with two SiPMs detecting light on both ends of the strips (later this configuration is referred to as BD counter).

Based on that, three series of test measurements were done:
\begin{enumerate}
\item Time resolution of Be-Be counters candidates;
\item Strip Efficiency of the BM@N Barrel Detector~(BD);
\item Time resolution of the BD counters.
\end{enumerate}
The results of the test measurements are presented in sections
~\ref{sec.Time_BEBE}, \ref{sec.Eff_BD} and \ref{sec.Time_BD}, respectively.

\section{Time resolution of Be-Be counters candidates}\label{sec.Time_BEBE}

\subsection{Experimental setup}

In this analysis, cosmic rays were used as a radiation source. For the Be-Be tested counters, we evaluate three types of ultra-fast plastic scintillators: BC404, BC422, and BC422Q~(1.0\% benzophenone) from Saint-Gobain Crystals.

\begin{figure}[t]
\includegraphics[width=1.\textwidth]{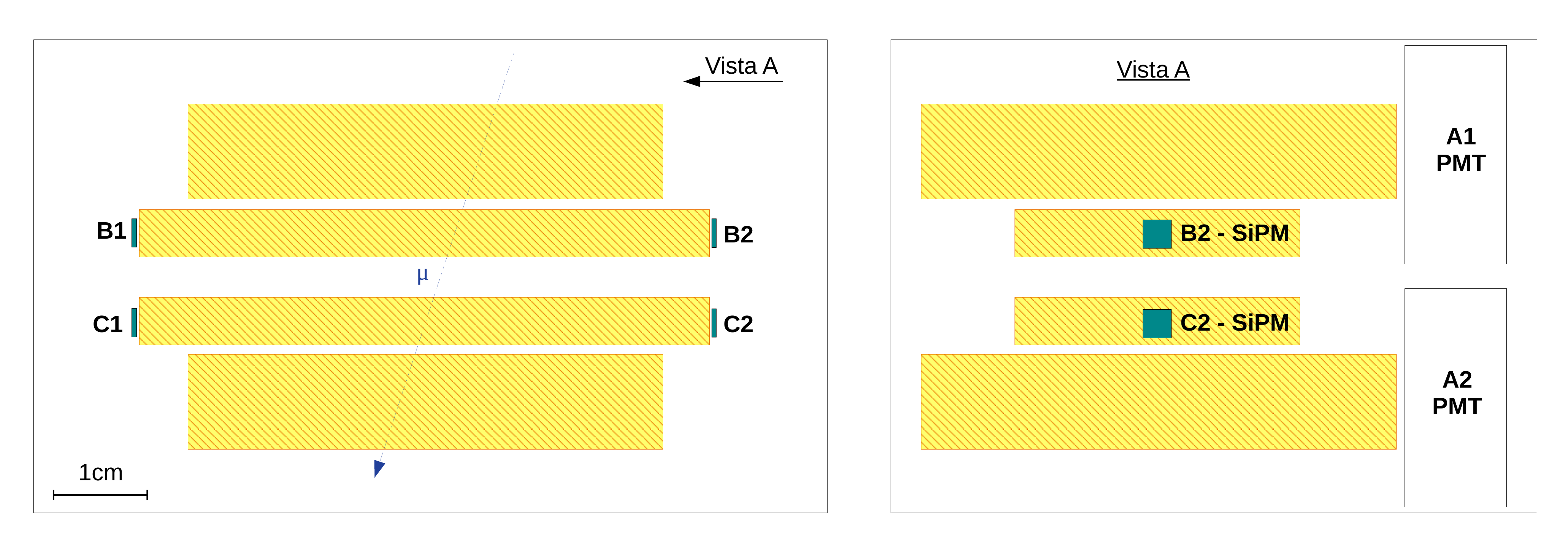}
\caption{Schematic layout of the experimental setup of Be-Be counters candidates.}
\label{BBsetup}
\end{figure}

The experimental setup is shown in fig.~\ref{BBsetup}. The tested counters are made of two SiPMs coupled to plastic scintillator BC-404 / 422/ 422Q wrapped in Al-mylar of $60\times30\times5mm^3$ and were placed between the trigger counters, which provided the start signal for data read-out. 

Four SiPMs Hamamatsu S13360-3050CS were used for the tests. The breakdown voltage is $54.1~V$. The SiPMs were placed at the ends of the strips with~(wg) and without~(wo) optical grease~(Saint-Gobain crystals BC-630). Because the SiPMs have $3\times3~mm^2$ sensitive area, while the cross section of the strip is $30\times5~mm^2$, each of the SIPMs cannot collect more than $6\%$ of incoming light.

\begin{figure}[b]
\begin{center}
\includegraphics[width=1.02\textwidth]{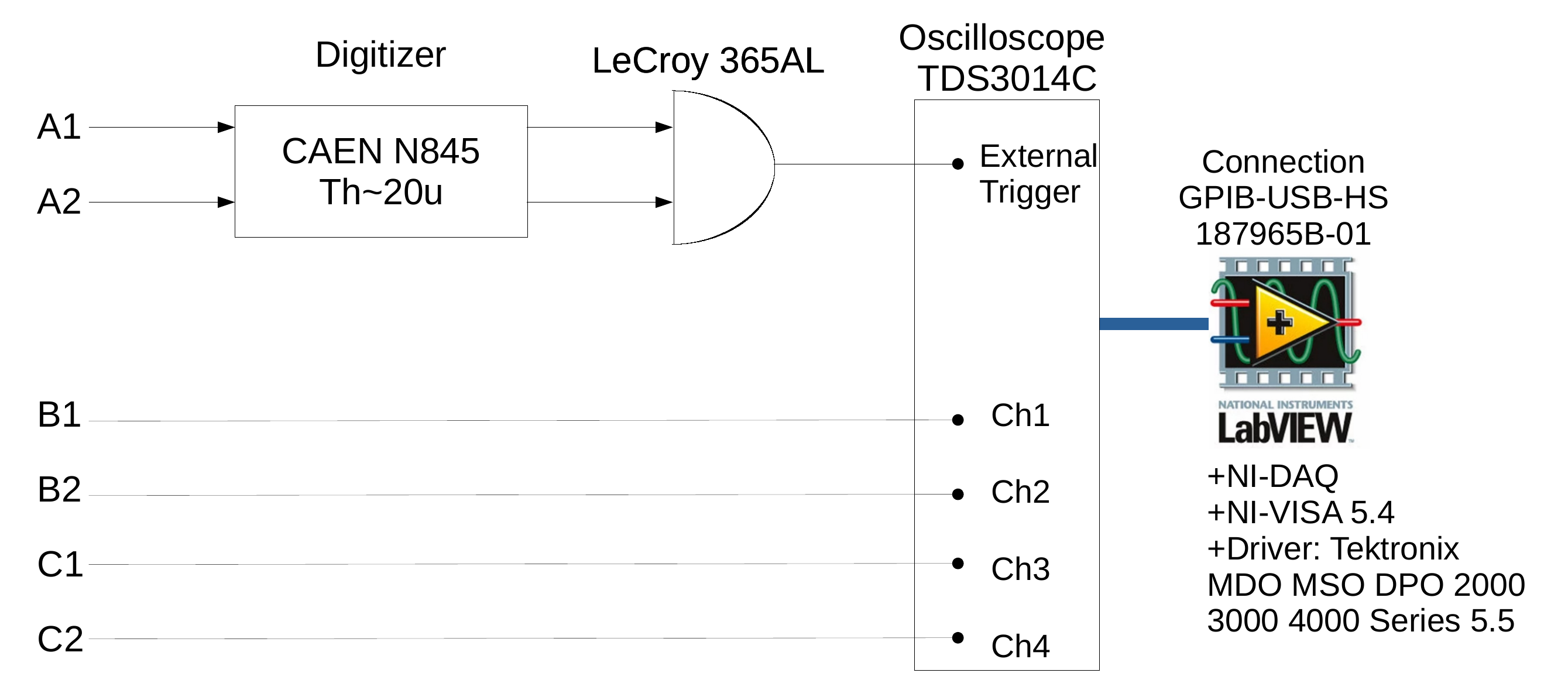}
\end{center}
\caption{Read-out electronics used for the Be-Be counters candidates.}
\label{BBread}
\end{figure}

The trigger counters were made of two scintillator bars~(BC404) of $50\times50\times10~mm^2$ and two PMTs~(Hamamatsu H5783). Light from the scintillators was detected by the PMTs at one end of the bar, and the coincidence of the signals from the two PMTs was the external trigger for the oscilloscope Tektronix TDS3014C. The read-out electronics is schematically shown in fig.~\ref{BBread}. 

\begin{table}[t]
\centering
\begin{tabular}{|c|c|c|c|c|c|c|}
\hline
Properties & \multicolumn{2}{|c|}{BC404} &\multicolumn{2}{|c|}{BC422} &\multicolumn{2}{|c|}{BC422Q} \\ \hline
~& wo & wg & wo & wg & wo & wg\\\hline\hline

Amplitude,~mV  & $14.4 \pm 0.2$&$19.5\pm0.2$ & $8.12 \pm 0.05$ & $14.8 \pm 0.1$ & $2.85 \pm 0.02$ & $4.60 \pm 0.03$ \\\hline
Rise time,~ns  & $16.1 \pm 0.8$ & $16.1 \pm 0.7$ & $16.2 \pm 0.6$ & $16.1 \pm 0.6$ & $15.9 \pm 0.9 $ & $15.9 \pm 0.8 $ \\\hline
Signal width,~ns & $50 \pm 2$ & $51 \pm 2$ & $51 \pm 3$ & $51 \pm 3$ & $50 \pm 2$ & $48.6 \pm 0.7$ \\\hline
\end{tabular}
\caption{Signal properties of the Be-Be candidates from the average of each SiPM pulse. The rise time was calculated with the difference between the times when the signal front crosses the $10\%$ and $90\%$ of the signal height. The signal width corresponds with the FWHM.}
\label{T_MeanS_BeBe}
\end{table}

The oscilloscope can digitize pulses in 4 input channels, which were used to sample pulses from the SiPMs of the test detectors.
The frequency of oscilloscope sampling was set at $100MHz$, which allowed to read-out $10,000$-bin-long waveform records with $0.2~ns$ time bin. Recorded data-sets were analyzed offline on the event-by-event basis. 

\begin{figure}[H]
\begin{center}
\includegraphics[width=0.44\textwidth]{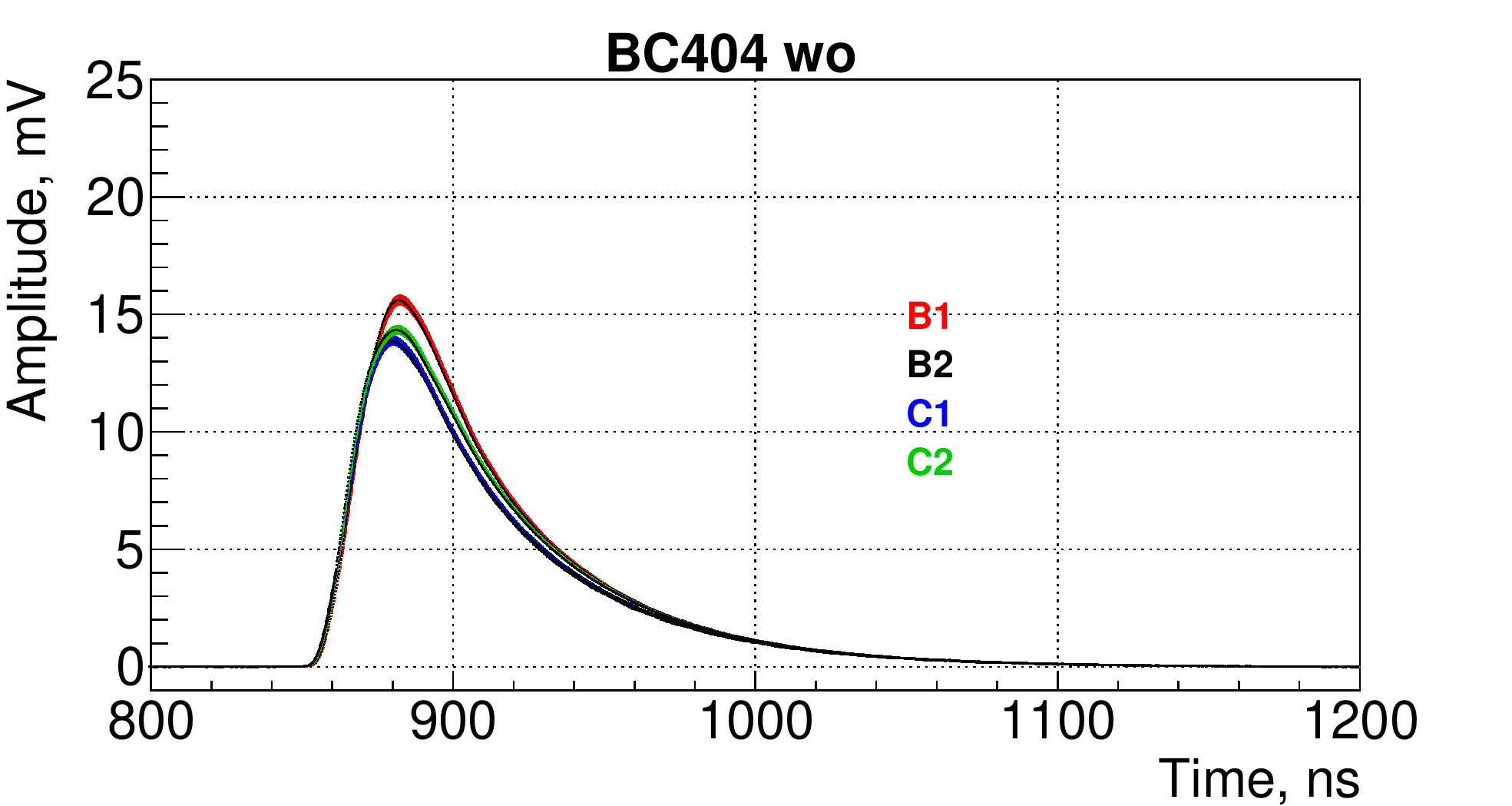}
\includegraphics[width=0.44\textwidth]{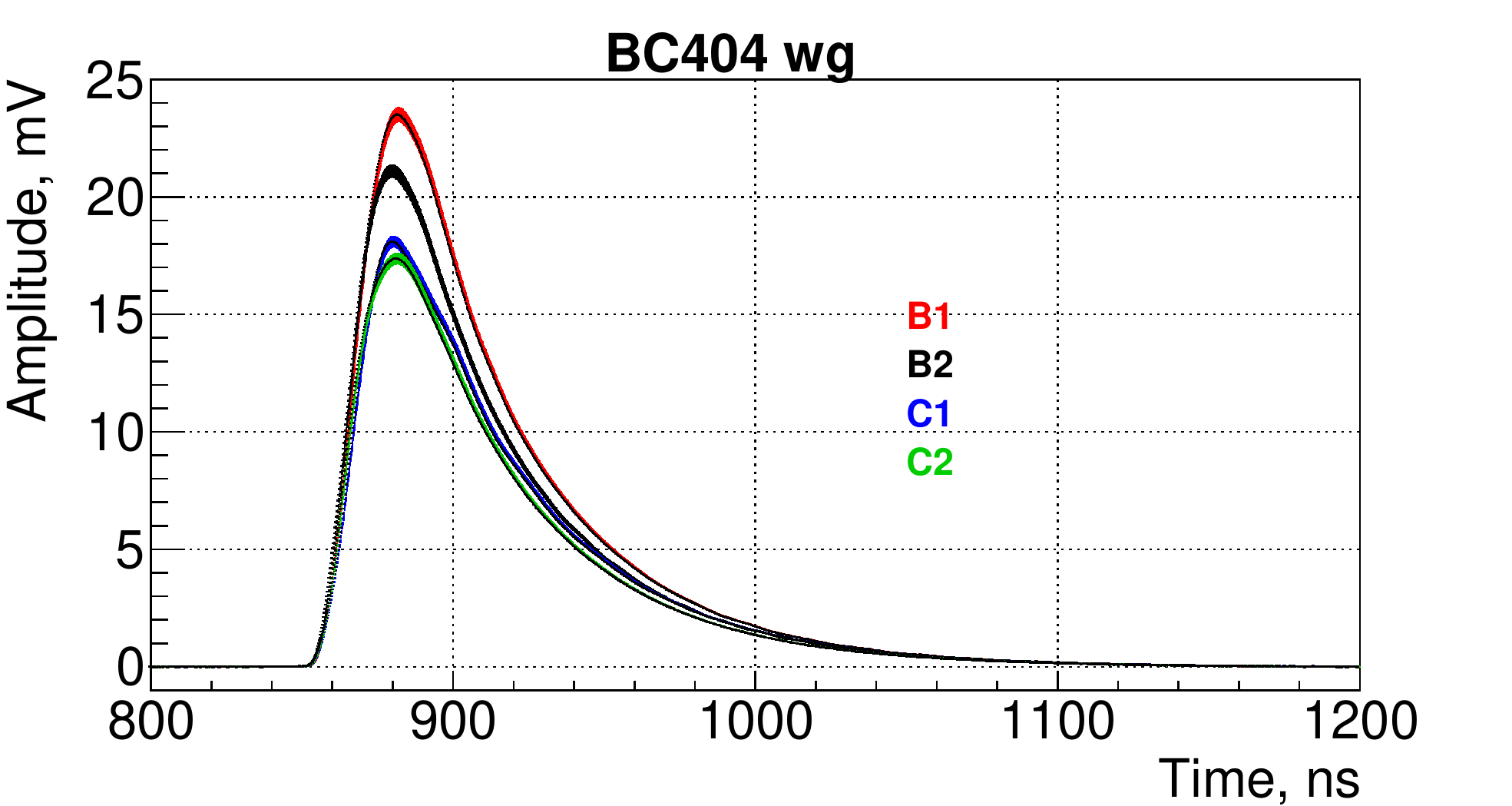}\\
\includegraphics[width=0.44\textwidth]{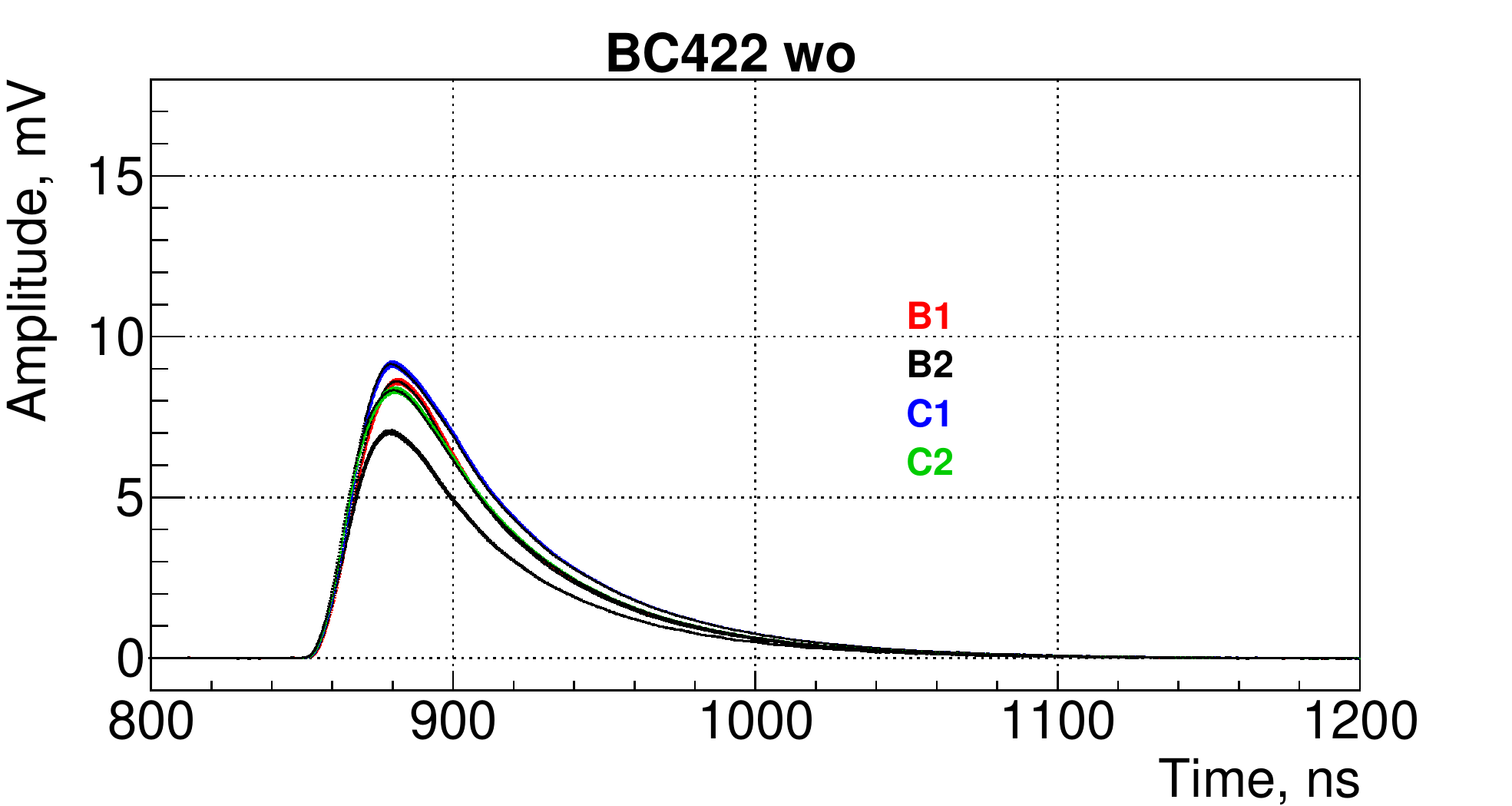}
\includegraphics[width=0.44\textwidth]{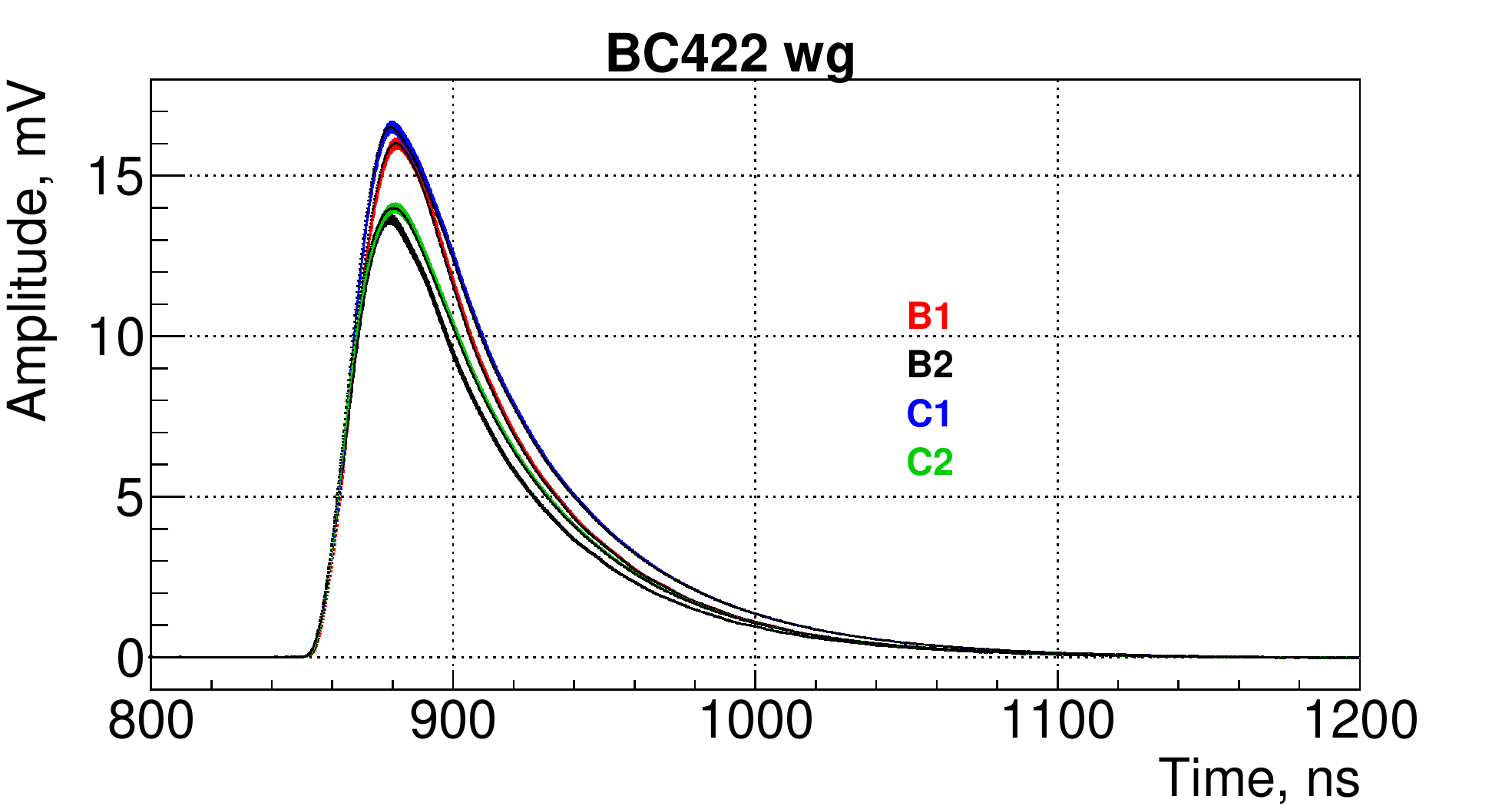}\\
\includegraphics[width=0.44\textwidth]{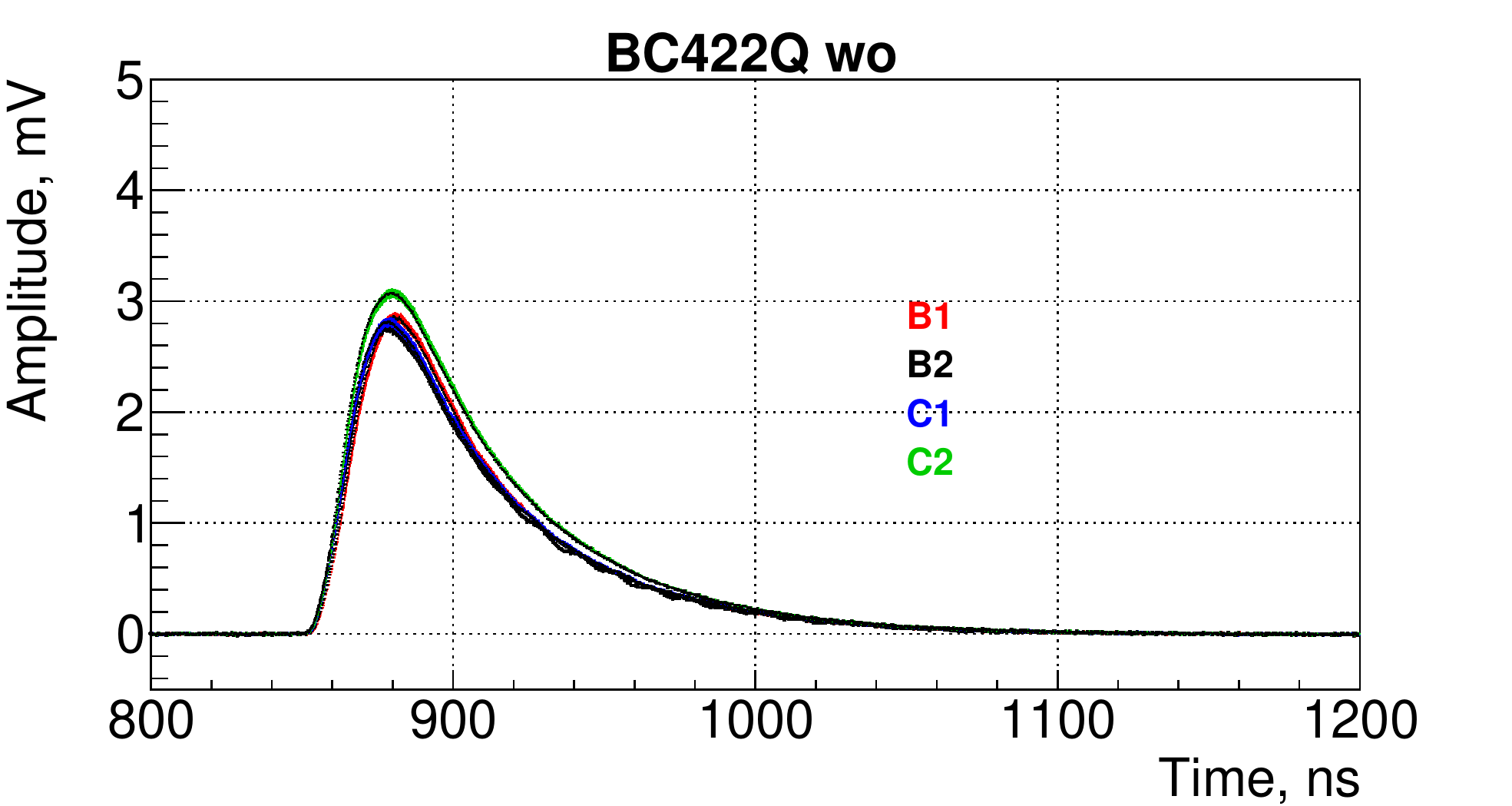}
\includegraphics[width=0.44\textwidth]{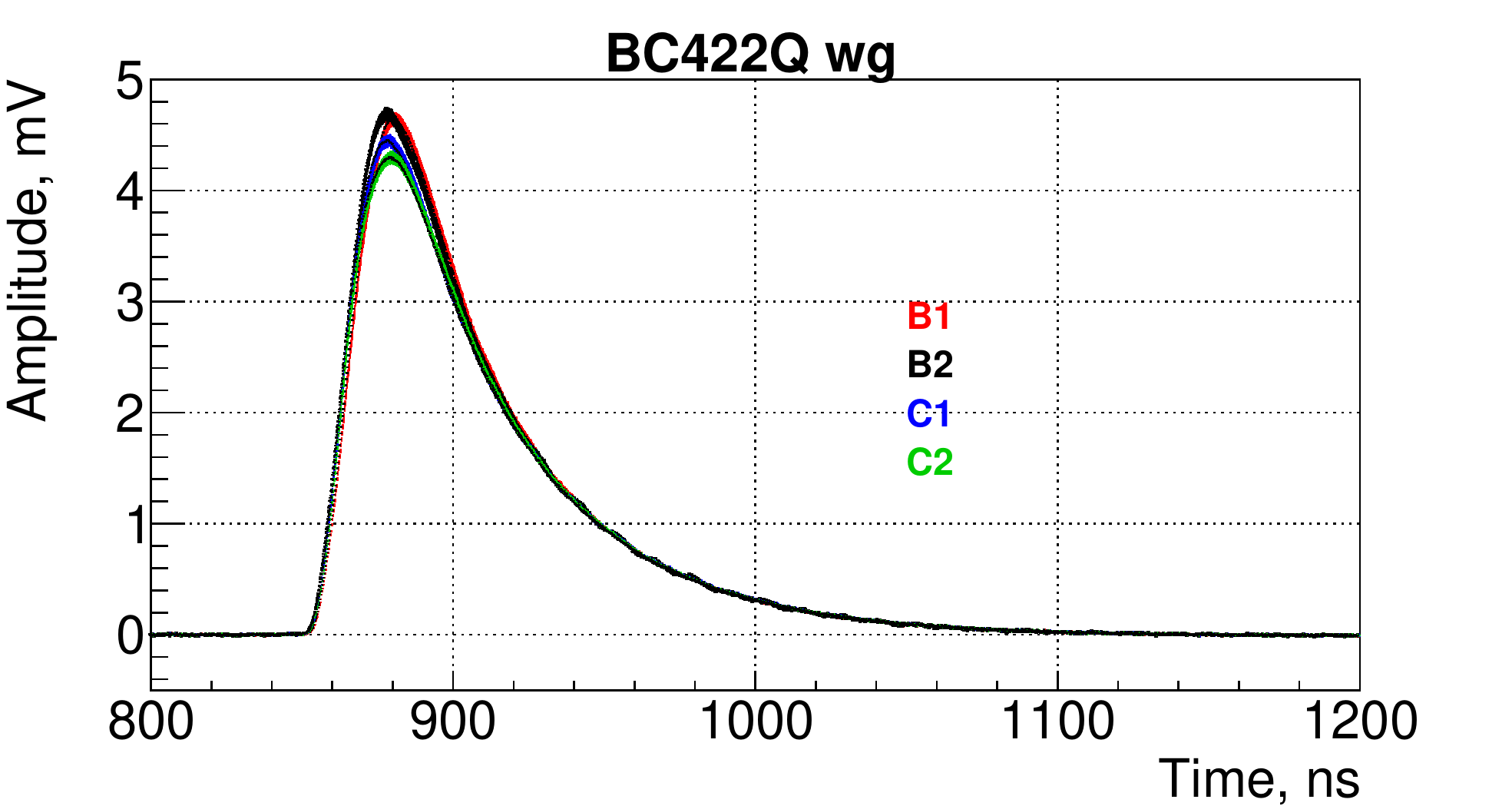}\\
\end{center}
\caption{Mean signal of each SiPM for BC404, BC422 and BC422Q form top to bottom, with~(right) and without~(left) optical grease.}
\label{MeanS_BeBe}
\end{figure}

The average of the signals coming from SiPMs B1, B2, C1 and C2 are showed in fig.~\ref{MeanS_BeBe}. When optical grease is applied, the pulse height increases due to improved light collection (for the BC-404 / 422 / 422Q the amplitude increased 40\% / 80\% / 100\%). The characteristic values from the average of the four mean signals are shown in table~\ref{T_MeanS_BeBe}. The rise time and the signal width did not show significant changes when the optical grease is applied.

\begin{figure}[b]
\begin{center}
\includegraphics[width=0.45\textwidth]{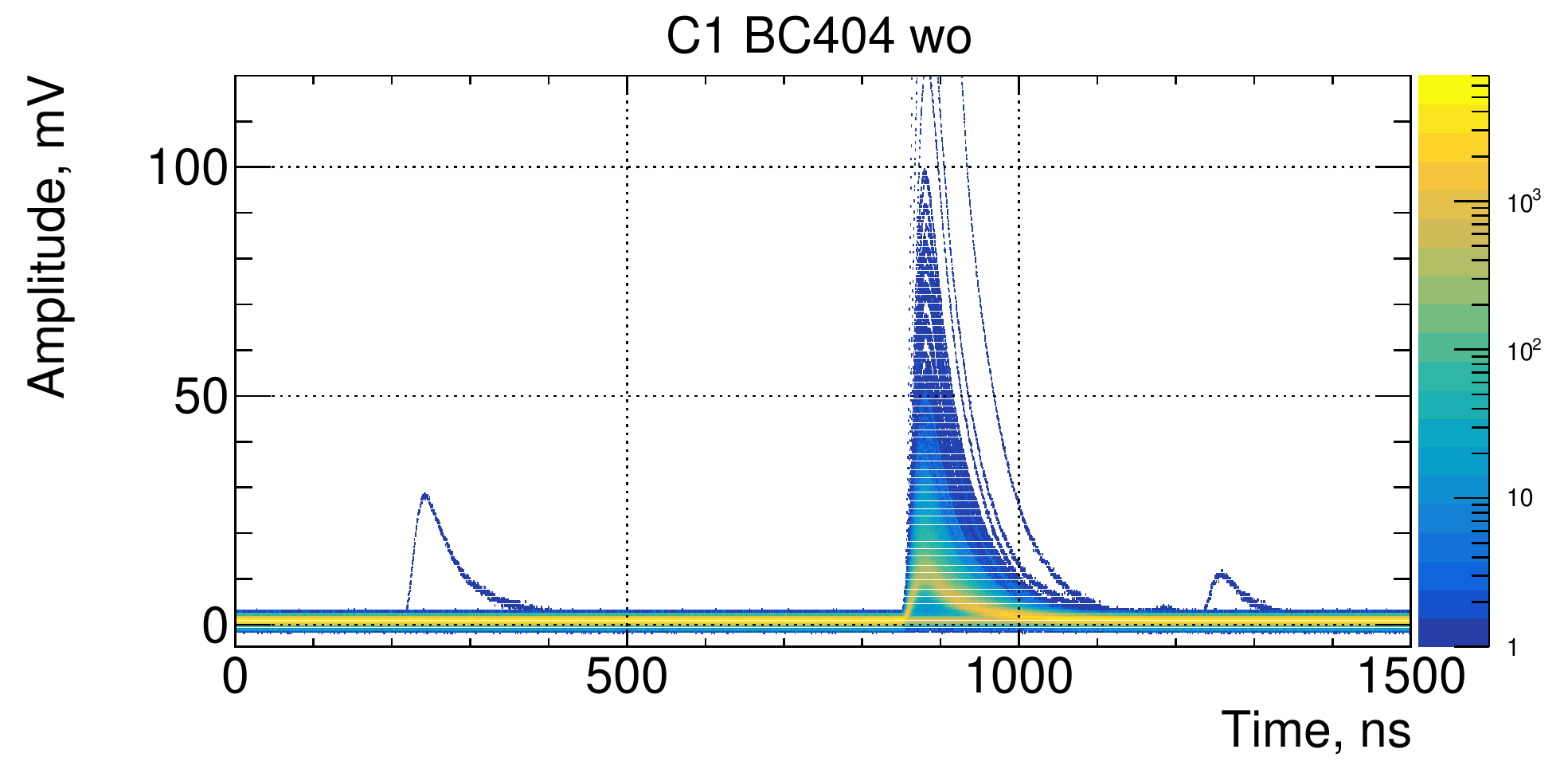}
\includegraphics[width=0.45\textwidth]{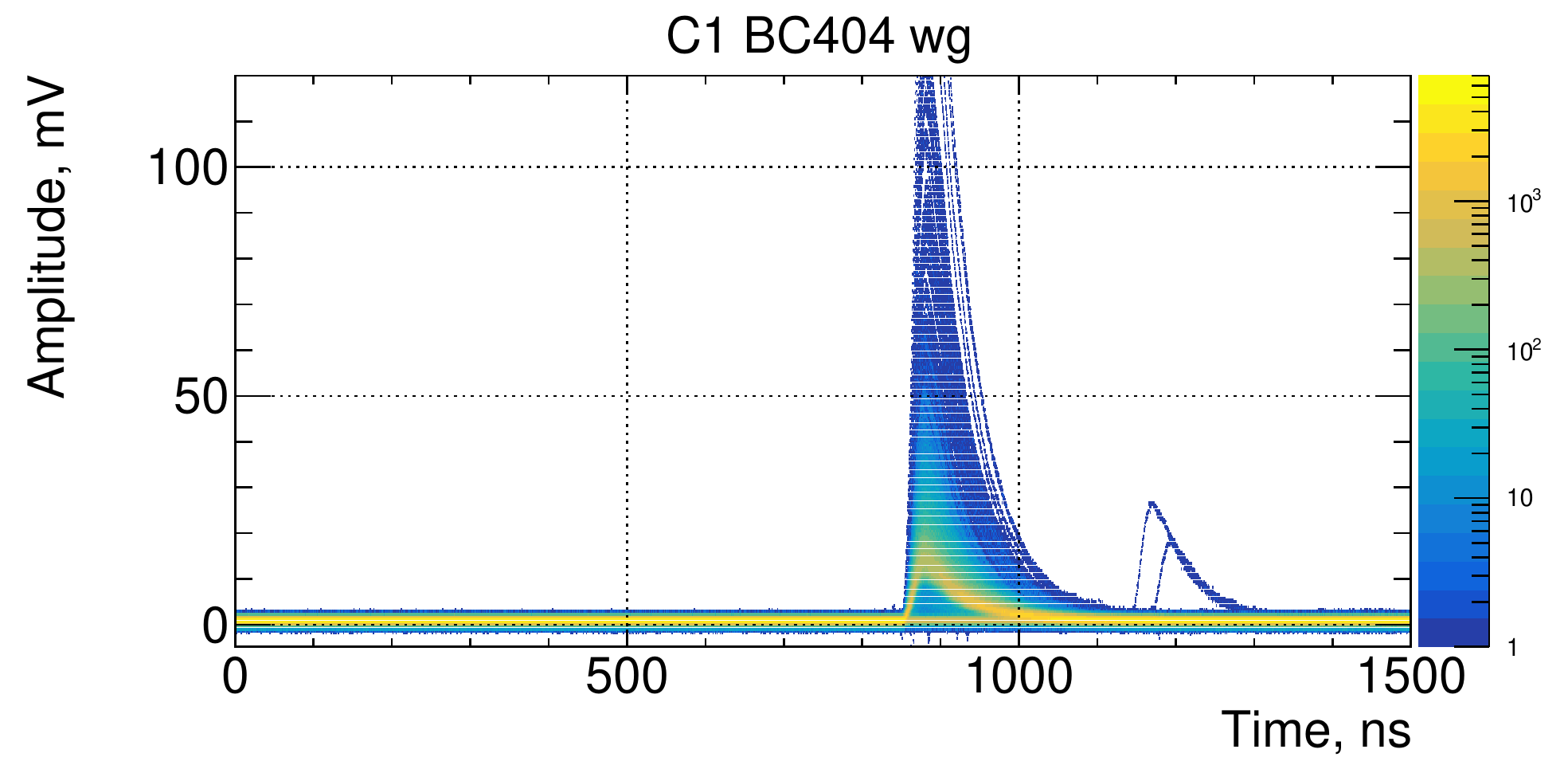}\\
\includegraphics[width=0.45\textwidth]{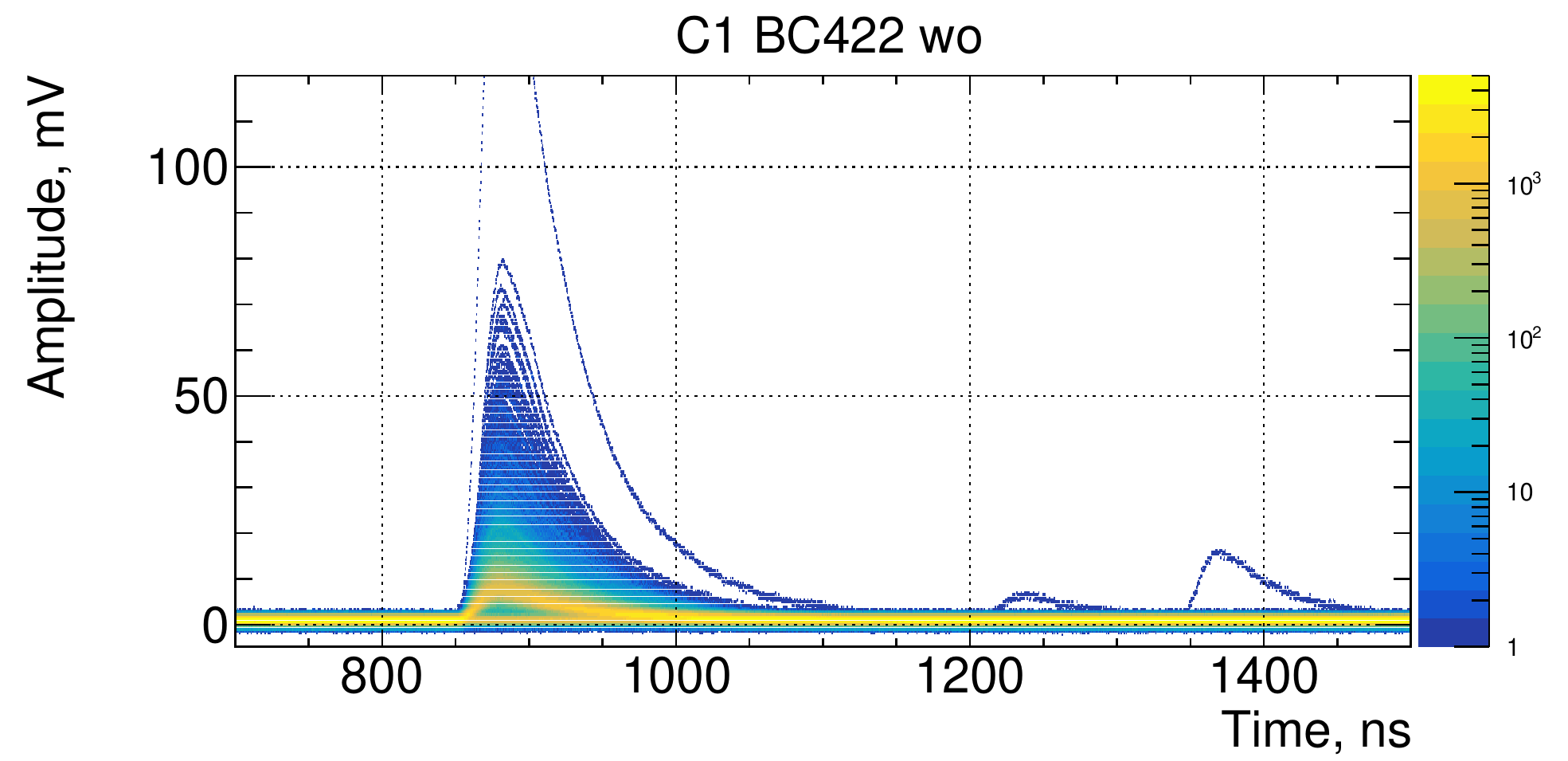}
\includegraphics[width=0.45\textwidth]{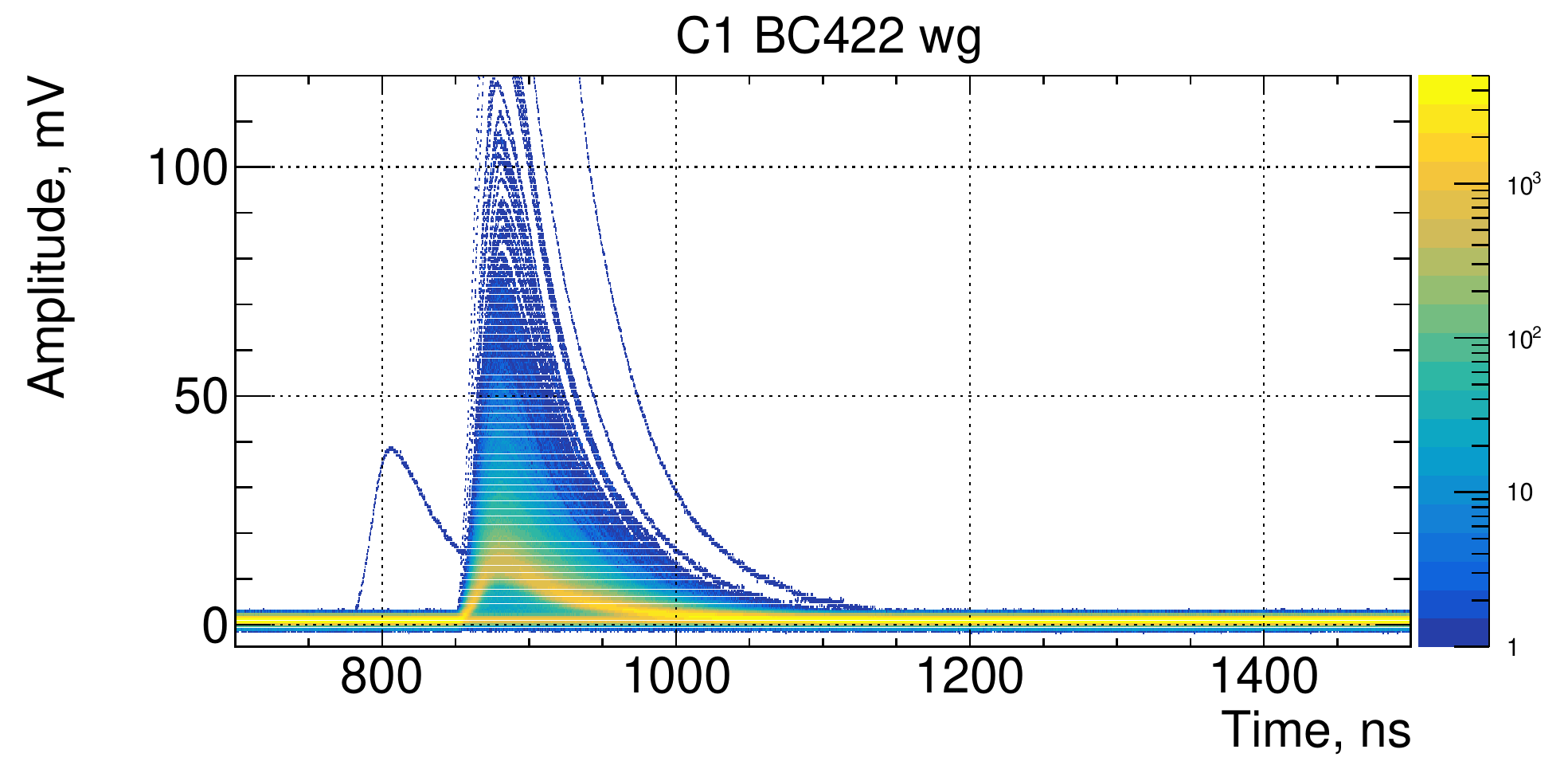}\\
\includegraphics[width=0.45\textwidth]{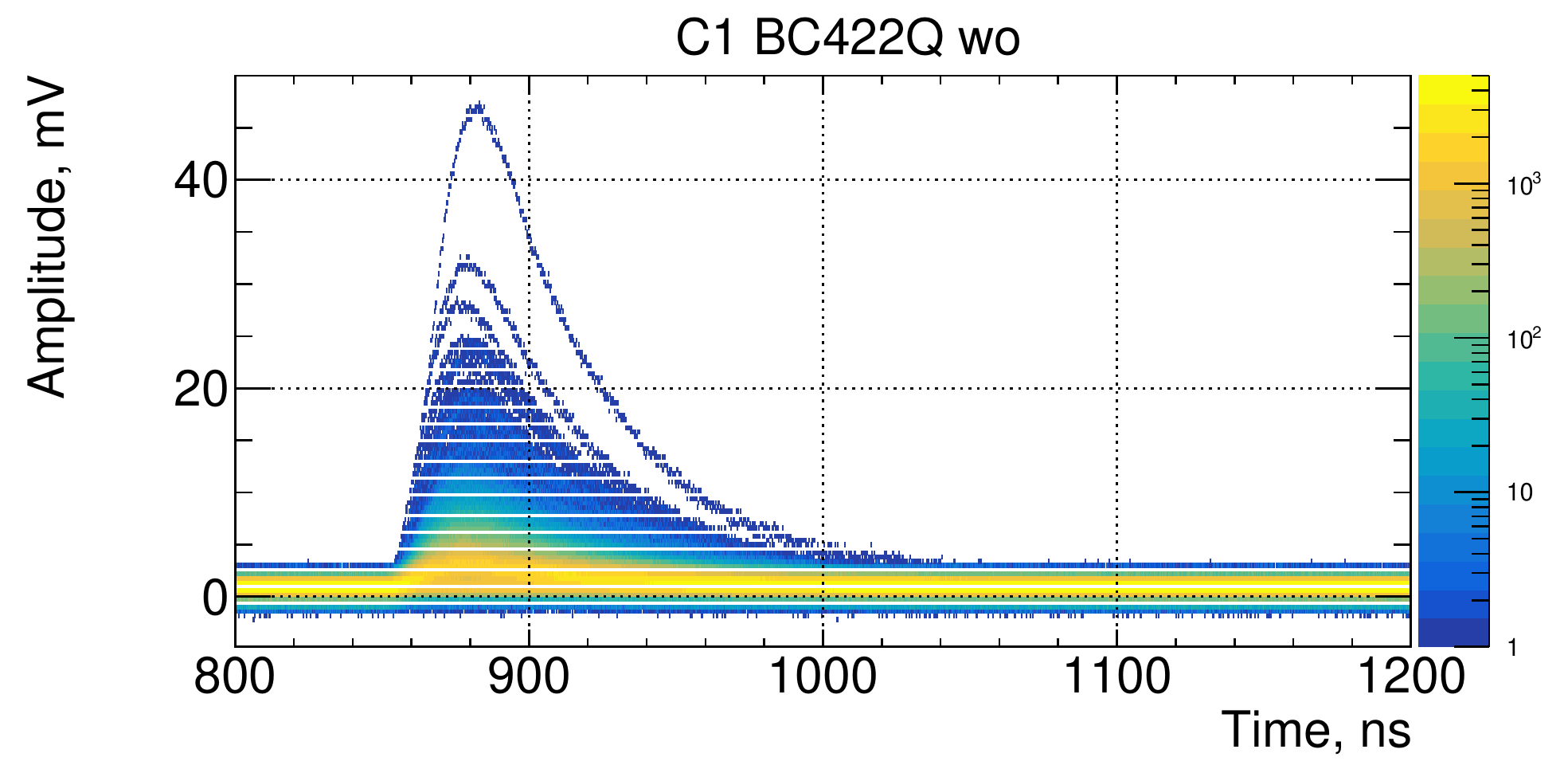}
\includegraphics[width=0.45\textwidth]{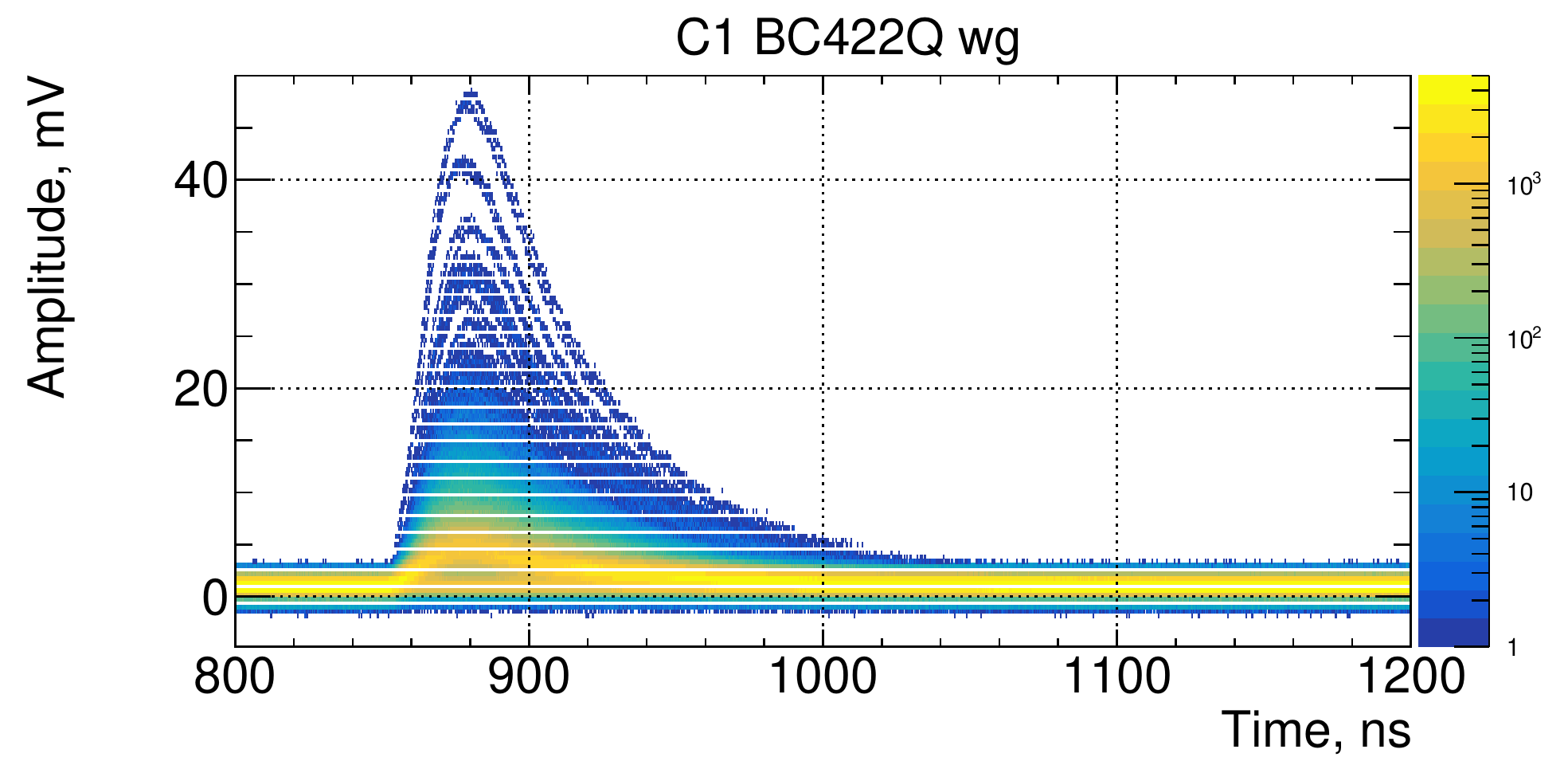}\\
\end{center}
\caption{Persistence of C1 for BC404, BC422, and BC422Q form top to bottom, with~(right) and without~(left) optical grease.}
\label{Per_BEBE}
\end{figure}

The signals from C1 with and without optical grease for the BC404, BC422, and BC422Q taken in the persistent mode of the oscilloscope are presented in fig.~\ref{Per_BEBE}. In all the cases, we can see a low noise and a few after pulsing events.

\begin{figure}[t]
\centering
\includegraphics[width=0.32\textwidth]{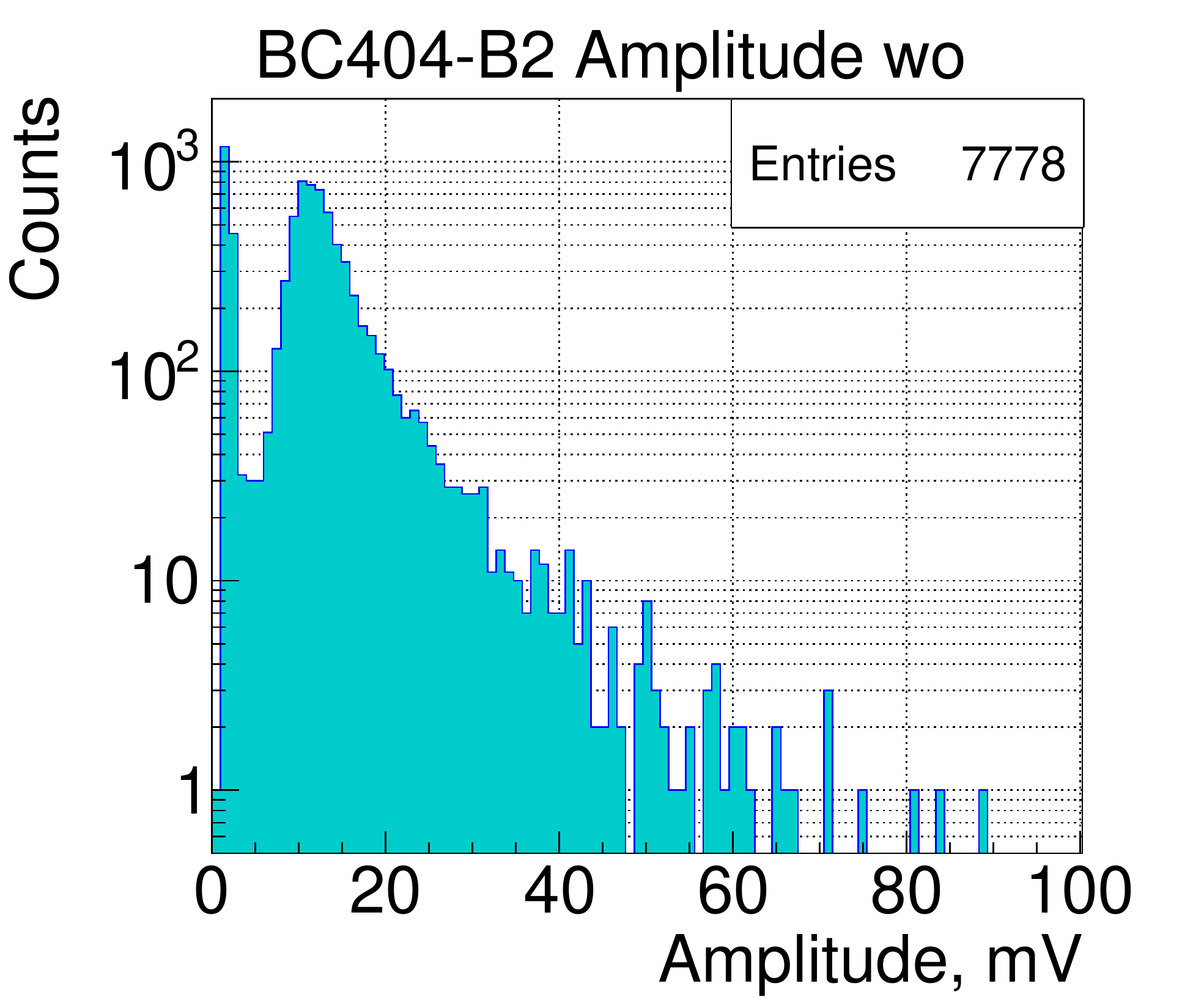}
\includegraphics[width=0.32\textwidth]{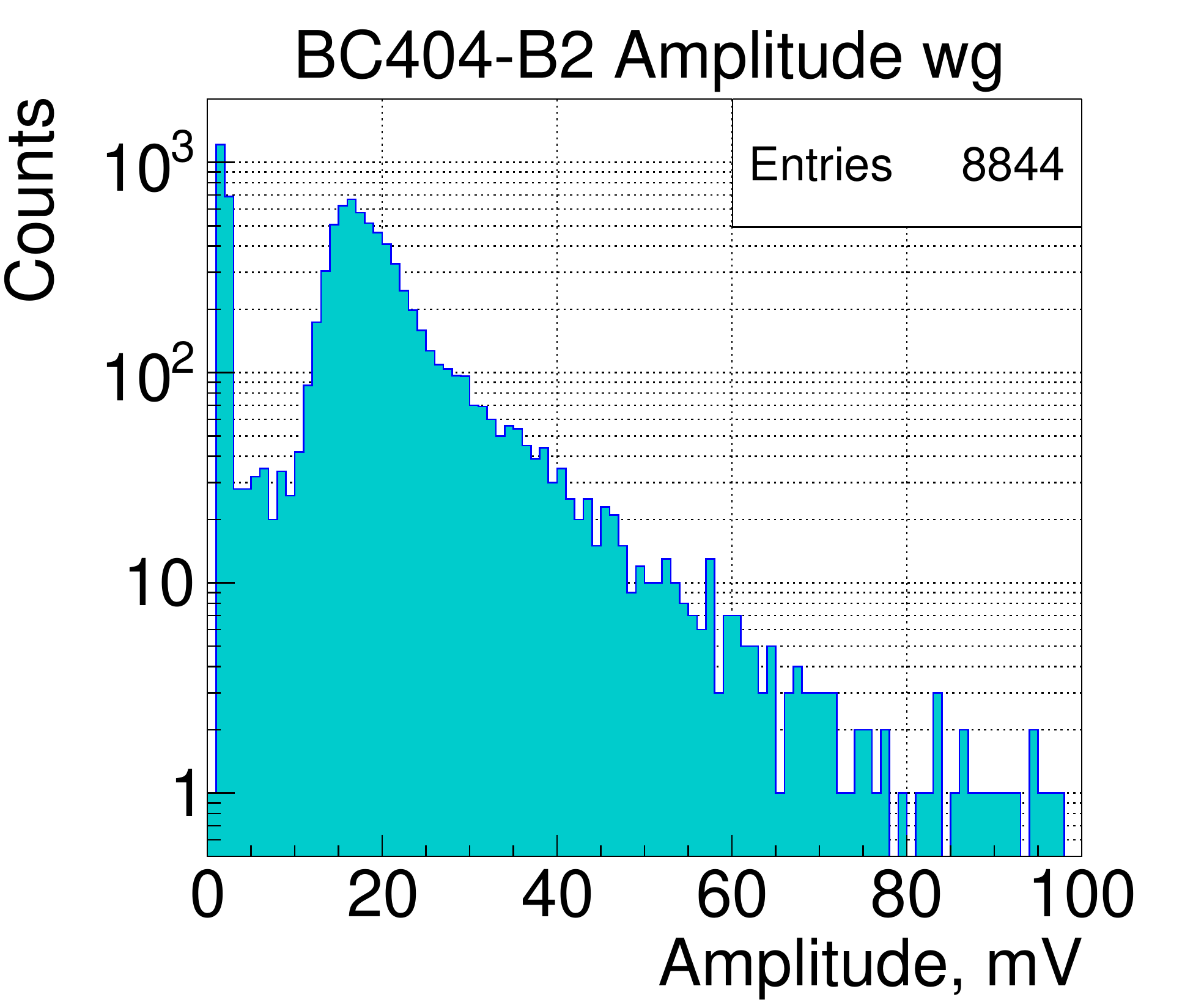}
\includegraphics[width=0.32\textwidth]{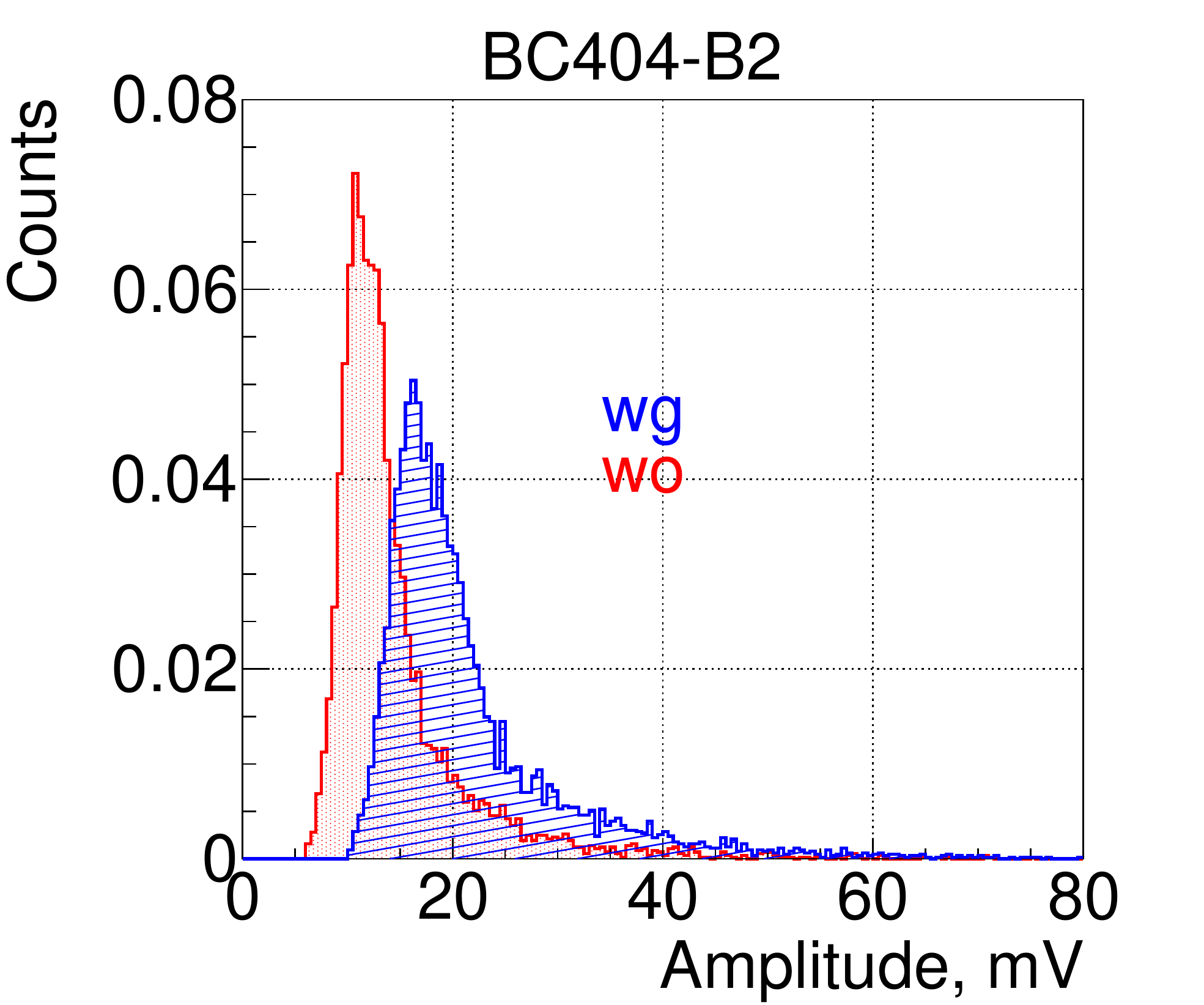}\\
\caption{Amplitude distributions of BC404 with~(center) and without~(left) optical grease for B2. The normalized amplitude distributions~(right plot) are showed after the selection criteria with~(blue) and without~(red) optical grease.}\vspace{1cm}
\label{Amp_404}
\end{figure}

\begin{figure}[H]
\centering
\includegraphics[width=0.325\textwidth]{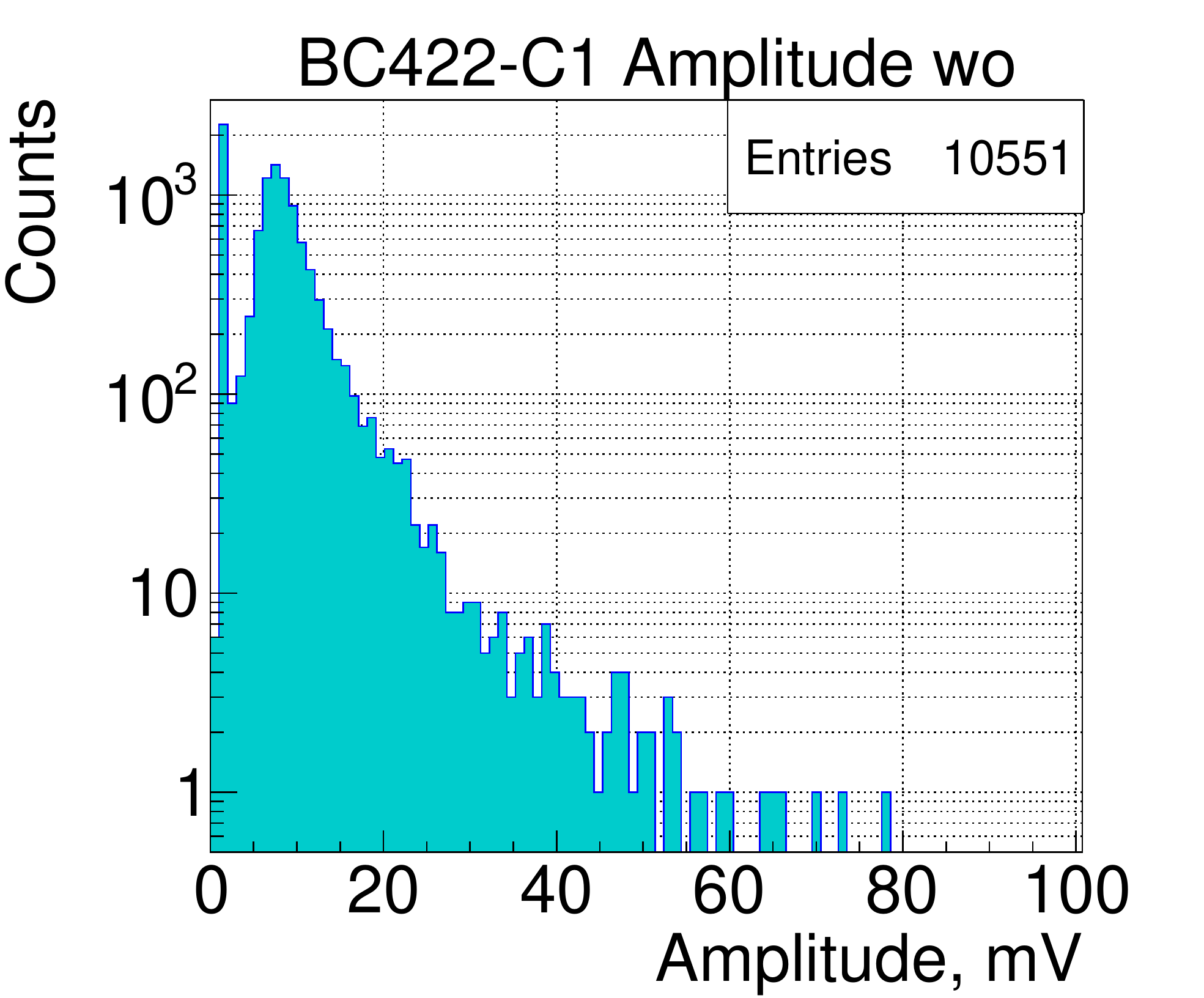}
\includegraphics[width=0.325\textwidth]{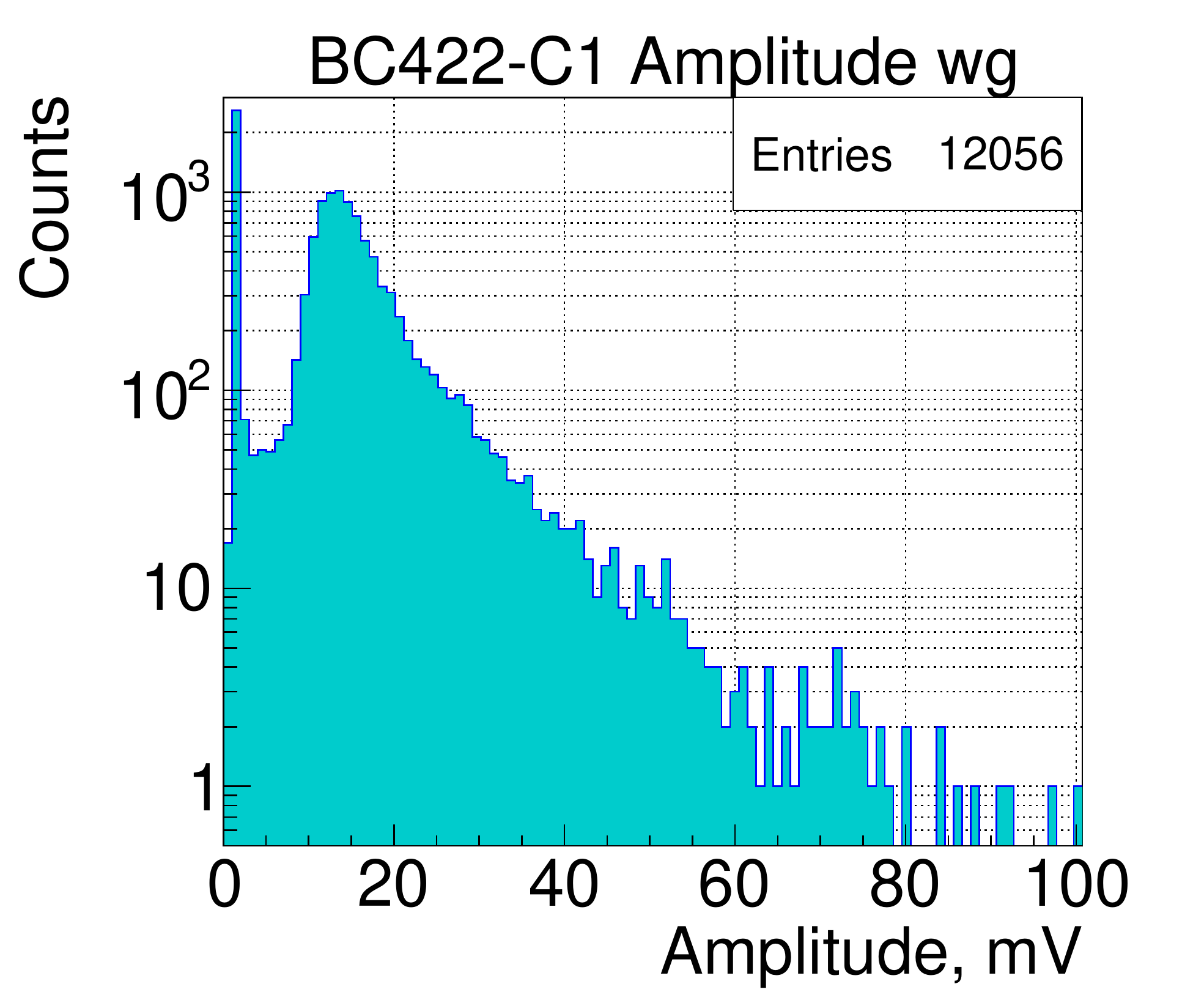}
\includegraphics[width=0.32\textwidth]{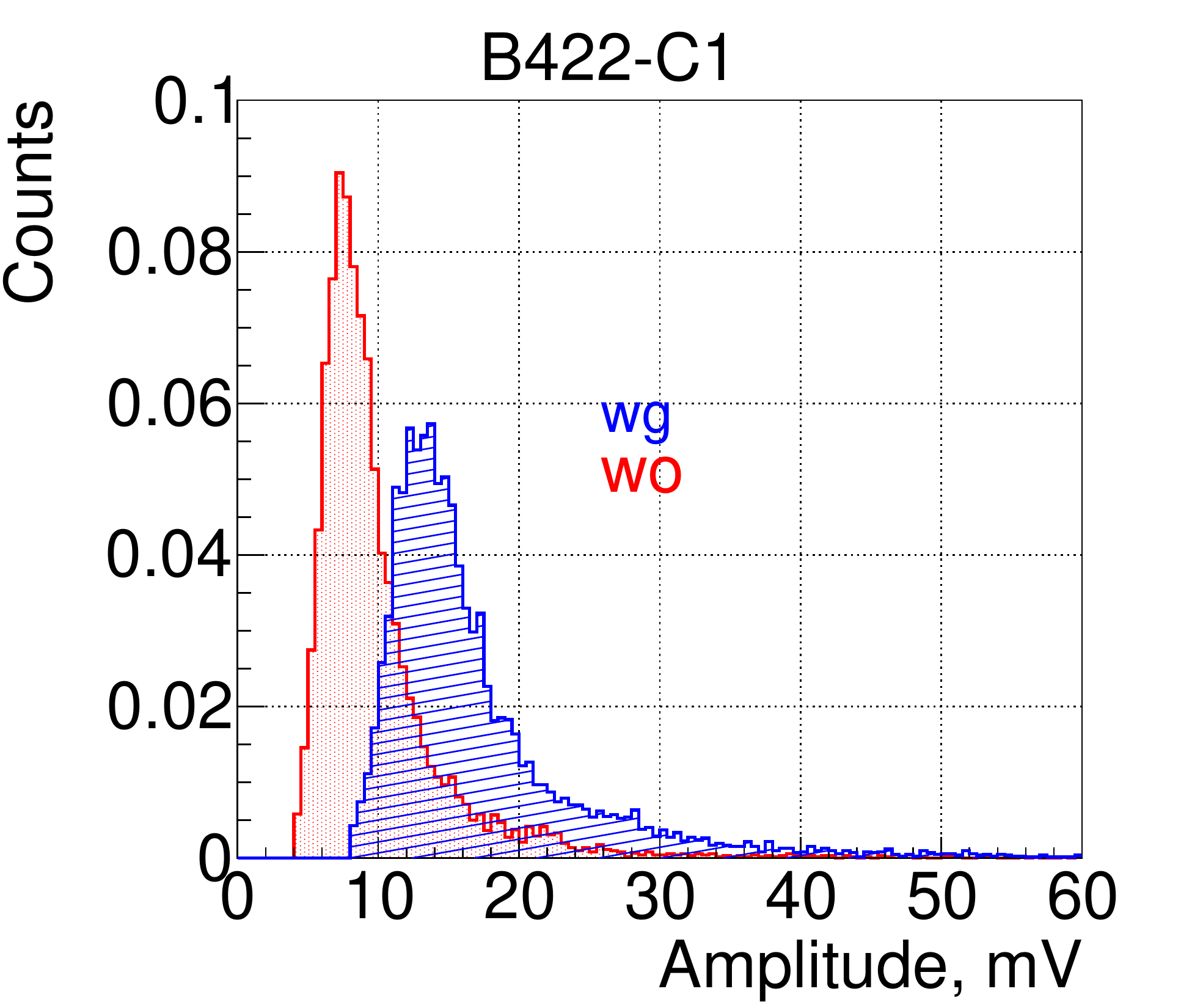}\\
\caption{The amplitude distributions of BC422 with~(center) and without~(left) optical grease for C1. The normalized amplitude distributions~(right plot) are showed after the selection criteria with~(blue) and without~(red) optical grease.}
\label{Amp_422}
\end{figure}

\begin{figure}[h]
\centering
\includegraphics[width=0.33\textwidth]{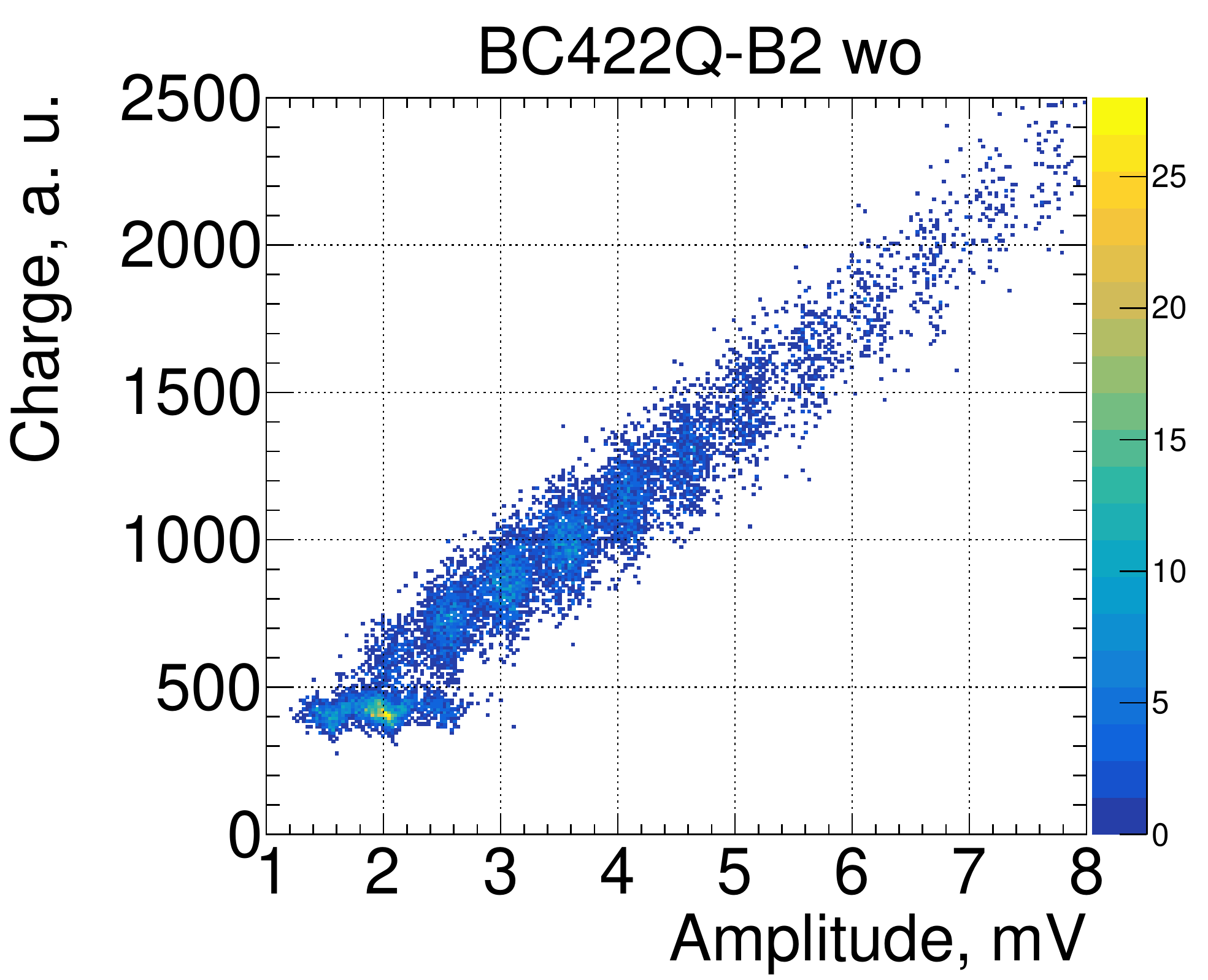}
\includegraphics[width=0.33\textwidth]{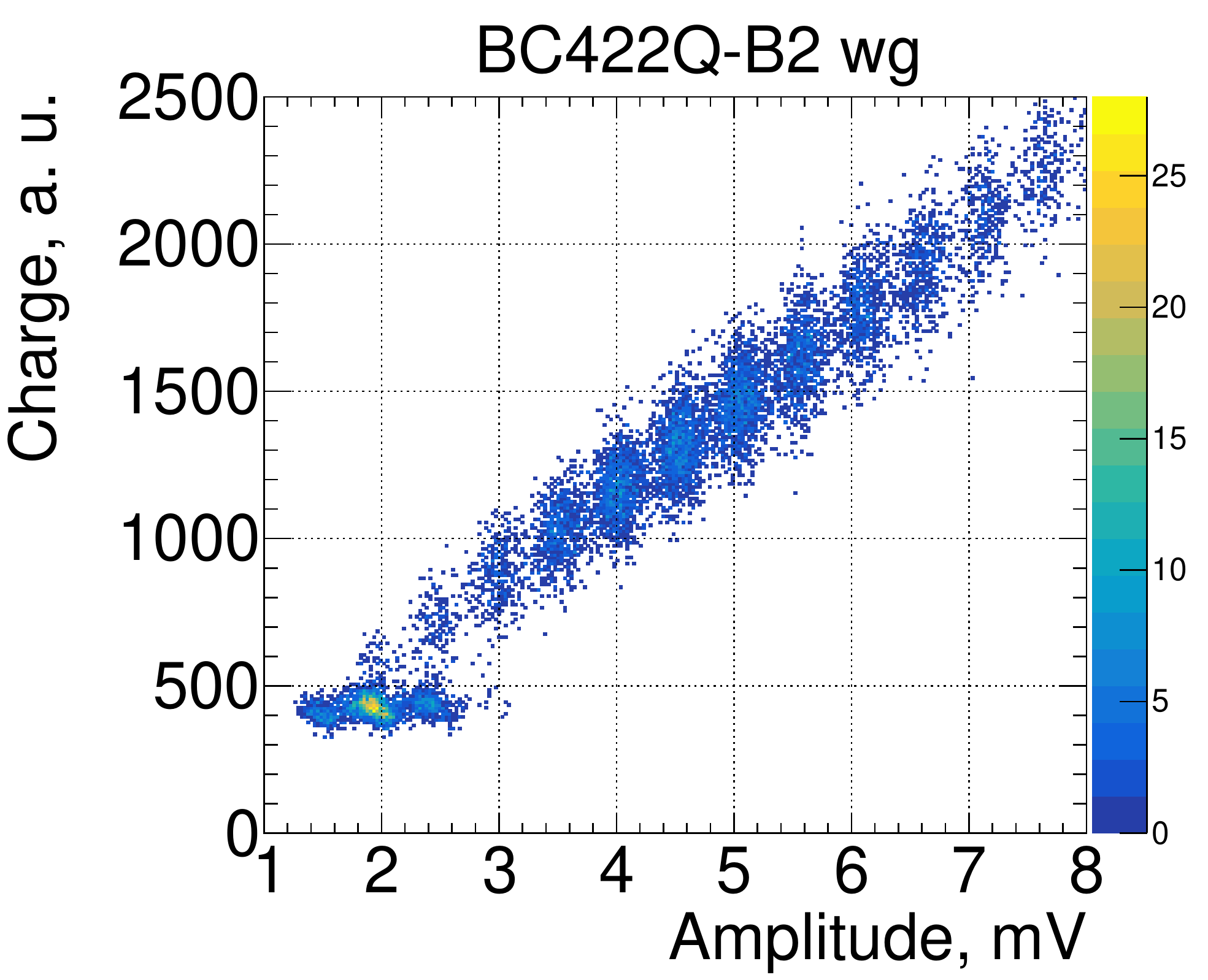}
\includegraphics[width=0.32\textwidth]{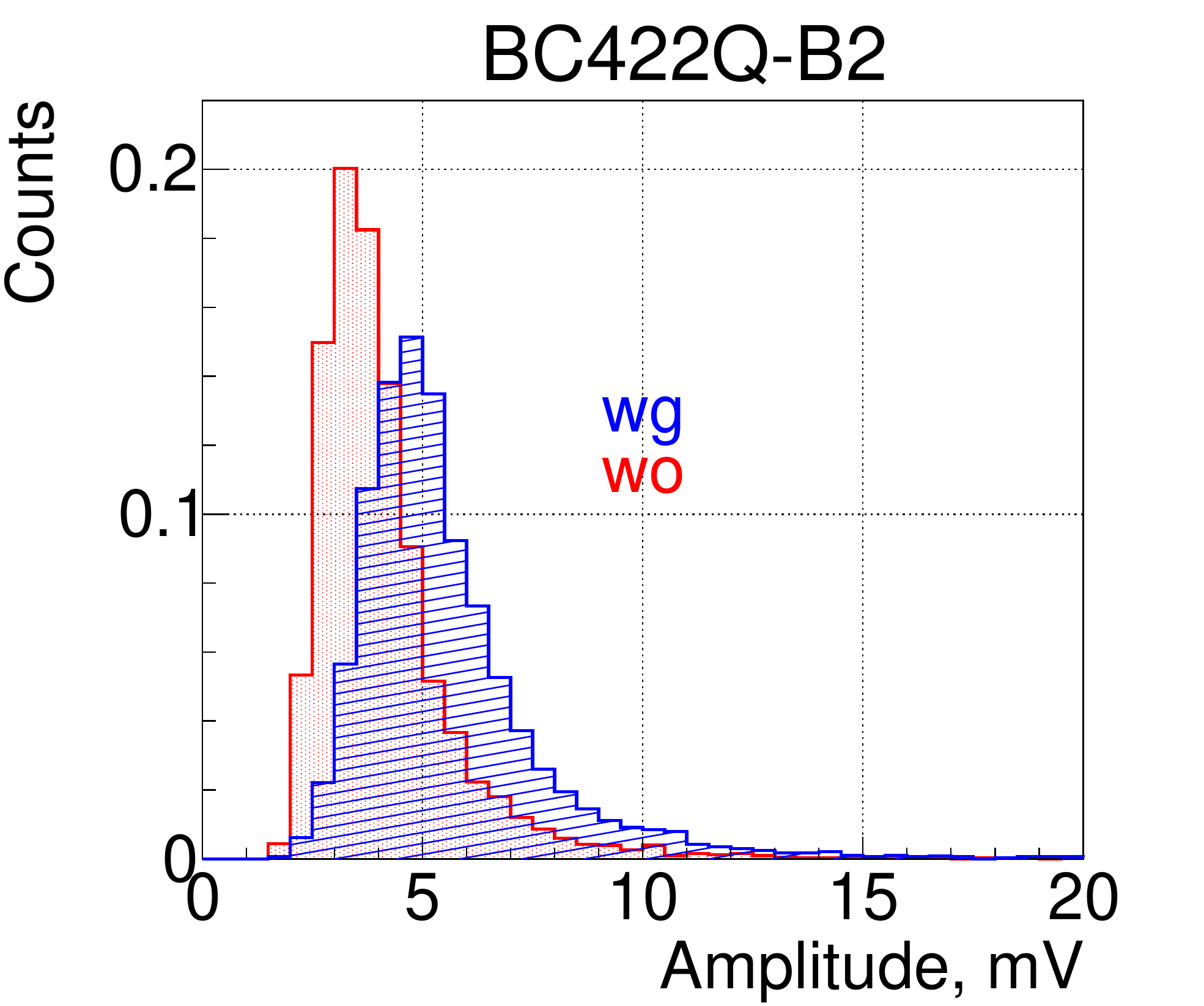}
\caption{The distribution of the charge vs. amplitude from the BC422Q signals with~(center) and without~(left) grease for B2. The events were chosen for the analysis corresponded to those having a charge~(a.~u.) value greater than 600. The normalized amplitude distributions~(right plot) are shown after the selection criteria with~(blue) and without~(red) optical grease.}
\label{Charge_Q}
\end{figure}

From the distributions of signals obtained in the persistent mode, the noise level was evaluated to establish a threshold limit for each SiPM. The threshold values for the selection of events for the following time resolution analysis were determined from the distributions obtained in the persistent mode and the distributions of signal amplitudes presented in fig.~\ref{Amp_404}-\ref{Charge_Q}.

As shown in fig.~\ref{Amp_404}~/~\ref{Amp_422} for BC-404~/~422 the distinction between the noise and the signal regions is clear, the noise is in the region less than $10~mV$~/~$8~mV$ with and  $7~mV$~/~$4~mV$ without optical grease; these were considered as the threshold values. For BC422Q, it was necessary to combine the persistence results with the charge vs. amplitude distribution, because the amplitude distribution did not show a clear distinction from the noise level. 

The charge vs. amplitude distribution for BC422Q is presented in fig.~\ref{Charge_Q}, and the noise corresponds to the region under 500 a.~u. of charge with and without optical grease, so, proper selection criteria correspond to a threshold level at 600 a. u. of charge.
Based on the above, for this analysis, events that satisfied the selection criteria were chosen.
Fig.~\ref{Amp_404}~/~\ref{Amp_422}~/~\ref{Charge_Q} shown the amplitudes after the selection criteria~(right plot) for BC-404~/~422~/~422Q, the distributions were normalized with the total number of selected events~(integral). Notice that there was a clear distinction when optical grease added.

\vspace{1cm}

\newpage
\subsection{Time resolution measurement for Be-Be counters candidates}\label{sec.TRBEBE}

\begin{figure}[b]
\centering
\includegraphics[width=0.85\textwidth]{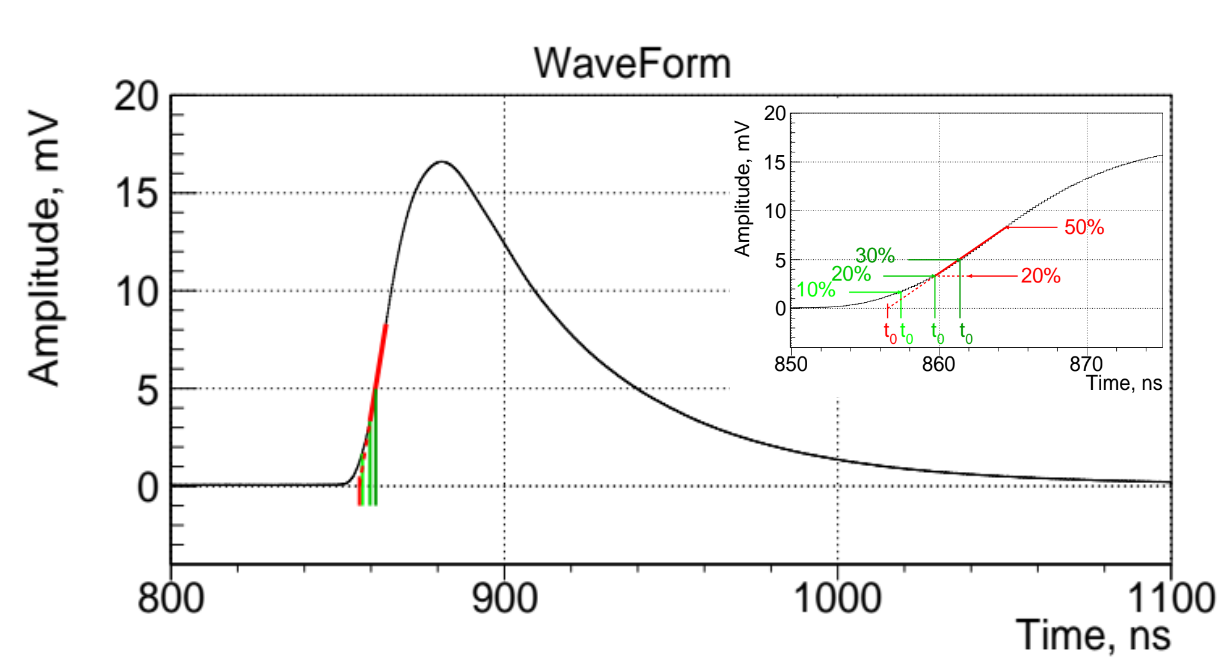}
\caption{Determination of the arrival time of the signal considering different pulse amplitude fractions~(\%)~(see text for details).}
\label{f_meth2}
\end{figure}

The time resolution can be affected by several factors: light output, light attenuation length and decay constant of the scintillator, the geometry, and efficiency of light collection, noise level and pulse shaping in the electronics (see, for example,~\cite{P2,Cattaneo,Stoykov}). 
In addition, the procedure chosen for the calculation of the arrival time of the signals may also influence the resulting resolution. In this study, we evaluate different methods which are illustrated in fig.~\ref{f_meth2}.
These methods represent some software variations of the constant fraction discriminator method~(CFD). 
In method M1, shown in red, we find two points in time, when the front of the input signal crosses $20\%$ and $50\%$ thresholds relative to the pulse maximum. Further, the region between these two points is fitted with a straight line. The intersection of the extrapolation of this line with the baseline of the waveform is set as the signal arrival time (t0).
In the methods M2, M3, and M4, shown in green, the t0 is defined as the point in time, when the front of the signal crosses the threshold of $10\%$, $20\%$ and $30\%$, respectively, relative to the pulse maximum.

\begin{figure}[t]
\centering
\includegraphics[width=0.49\textwidth]{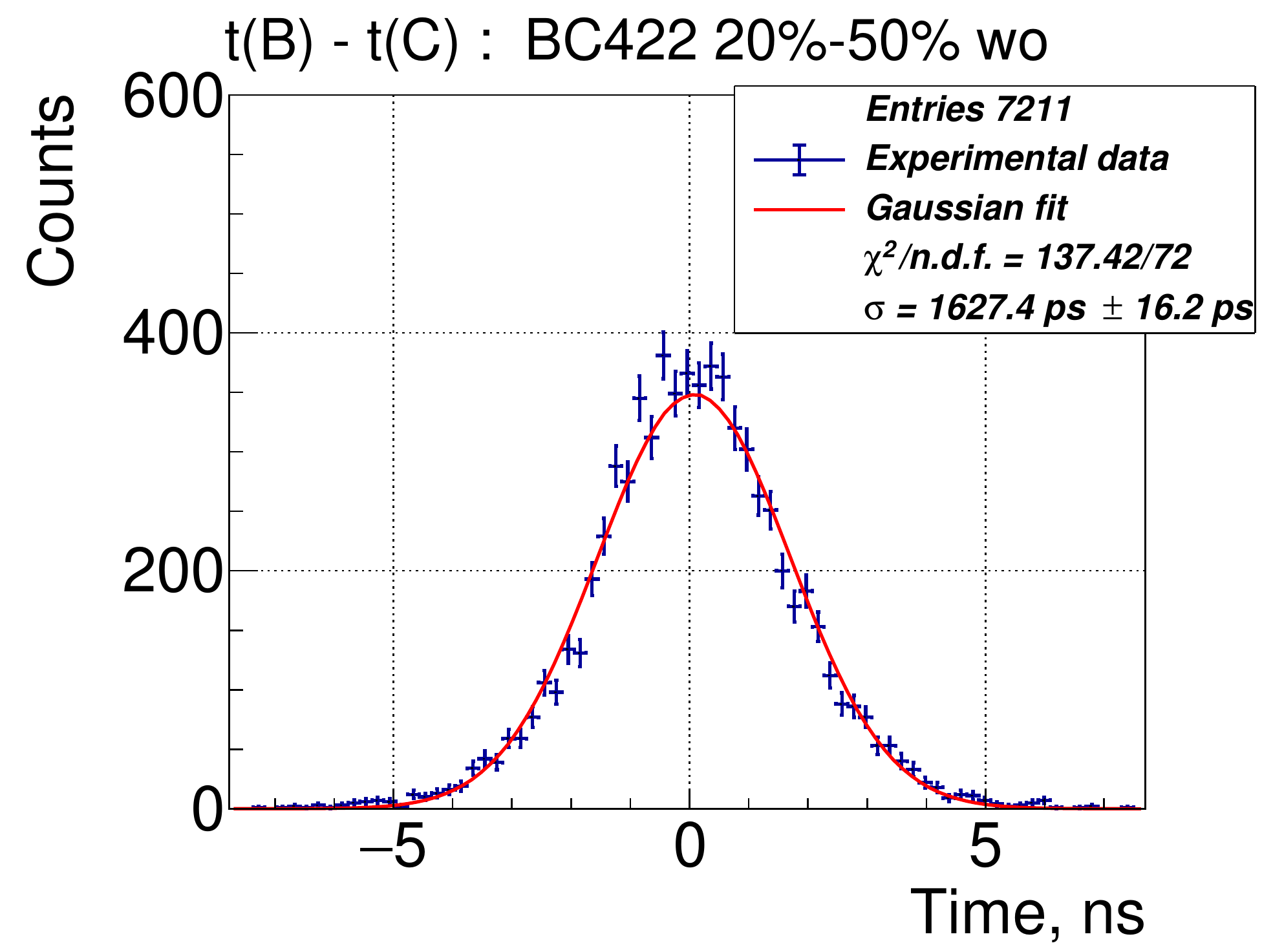}
\includegraphics[width=0.49\textwidth]{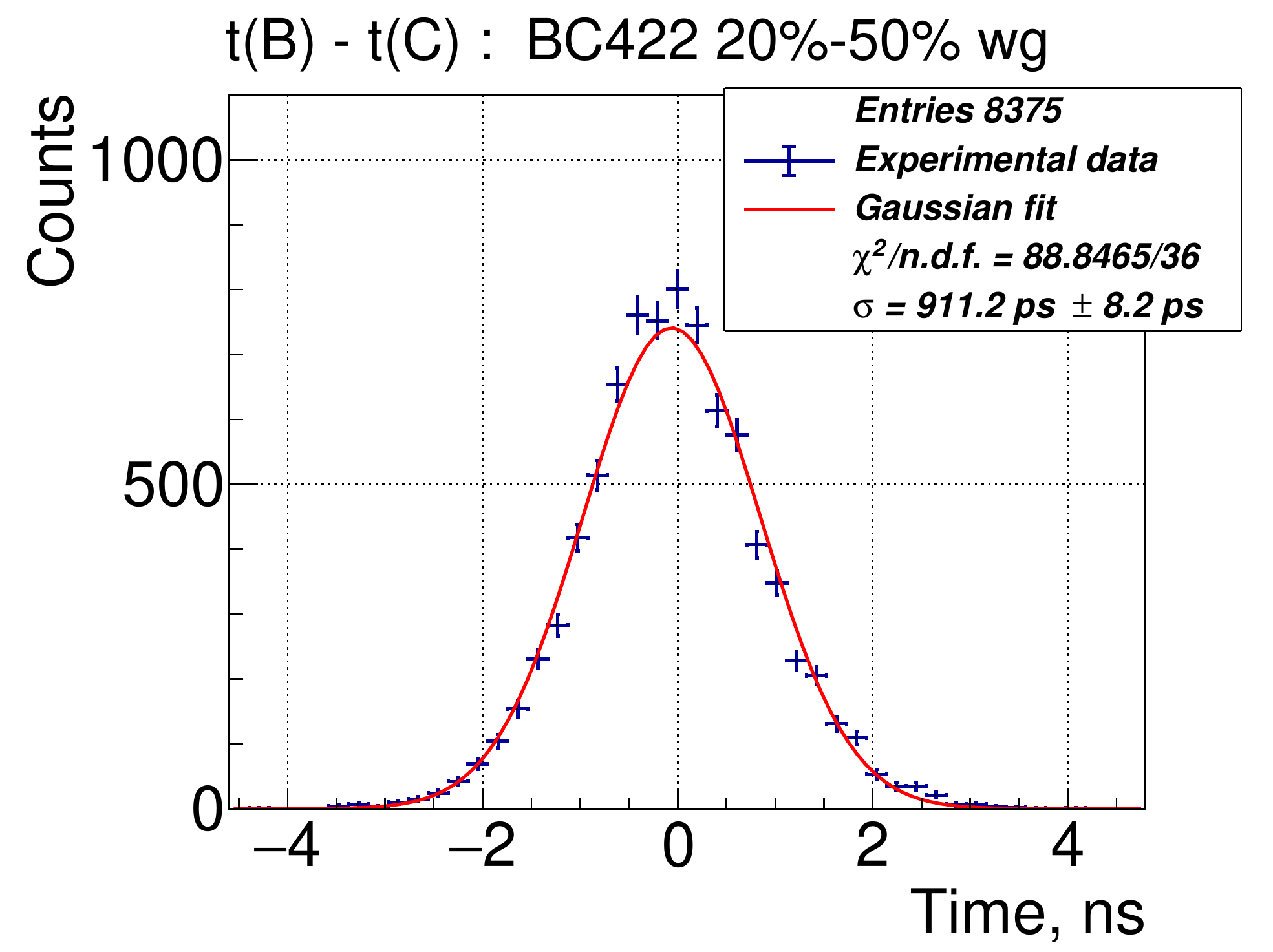}
\caption{Time difference distributions of BC422 with~(right) and  without~(left) optical grease obtained using the M1 method.}
\label{TR_distributions}
\end{figure}

The cosmic rays impact times in the detectors B and C were calculated as $t_B = \frac{1}{2} (t0_{B1}+t0_{B2})$ and $t_C =\frac{1}{2}  (t0_{C1}+t0_{C2})$, respectively. Then, the difference between these times ($t(B)-t(C)$) was obtained using M1 method showed in fig.~\ref{TR_distributions} for  BC422, where the standard deviation corresponds to $\sigma_{BC}$. Similar distributions were done for the other plastics and methods. If it is taken into account that both counters B and C have the same time resolution since both are identical and work under the same conditions, then, ~$\sigma_B=\sigma_C=\frac{\sigma_{BC}}{\sqrt{2}}$. Table~\ref{times_BeBe} shows the time resolution of the counters obtained by each method.

\begin{figure}[tbh]
\centering
\includegraphics[width=0.49\textwidth]{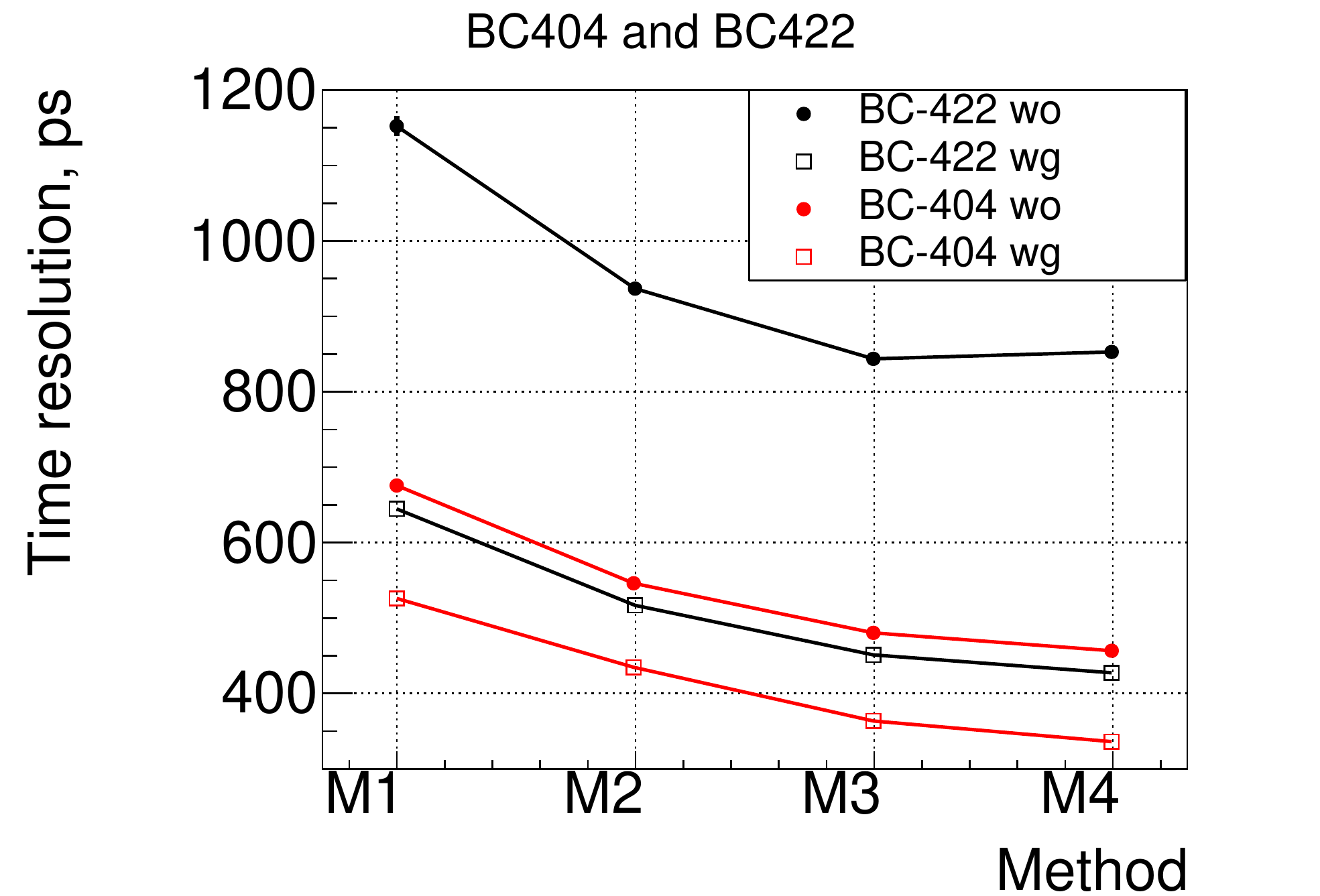}
\includegraphics[width=0.49\textwidth]{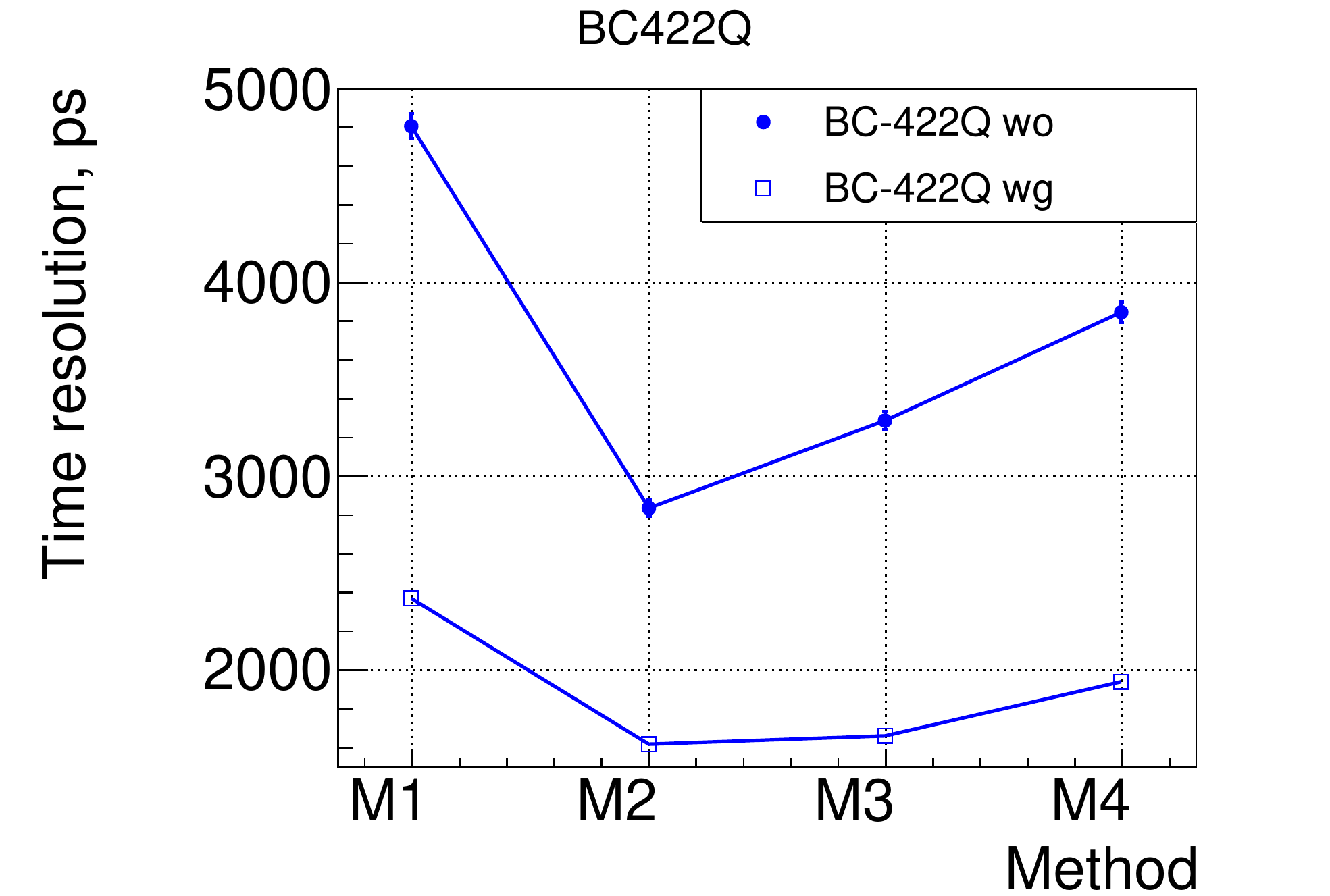}
\caption{Time resolution of each counter with~(wg) and without~(wo) optical grease obtained by the methods  M1~(20\%-50\%), M2~(10\%), M3~(20\%) and M4~(30\%).}
\label{TR_Abs}
\end{figure}

Fig.~\ref{TR_Abs} shows strong dependency of the time resolution on the method chosen to determine the arrival time of the signal, the difference of the values reach up to $40\%$. Independently of the method, in all the cases, when the optical grease is applied, the time resolution is improved. From these results, the best time resolution obtained in this analysis were: M4 for BC404, M4 / M3 for BC422 wg / wo, and M2 for BC422Q.

\begin{table}[h]
\centering
\begin{tabular}{|c|c|c|c|c|c|c|}
\hline
{Method} & \multicolumn{2}{|c|}{BC404~(ps)} & \multicolumn{2}{|c|}{BC422~(ps)} & \multicolumn{2}{|c|}{BC422Q~(ps)}\\\cline{2-7}
 CFD (\%)& wo & wg& wo & wg& wo & wg\\
\hline \hline
M1~(20-50) 	& $675 \pm 7$&$525\pm6$ & $1151\pm12$&$644\pm6$ & $4799\pm64$&$2363\pm28$ \\\hline
M2~(10) 	& $545 \pm 6$&$432\pm4$ & $936\pm9$&$516\pm5$   & $2835\pm42$&$1616\pm18$ \\\hline
M3~(20) 	& $480 \pm 5$&$363\pm4$ & $842\pm8$&$451\pm4$   & $3281\pm48$&$1656\pm19$ \\\hline
M4~(30)		& $455 \pm 5$&$335\pm3$ & $852\pm9$&$427\pm4$   & $3842\pm52$&$1939\pm26$ \\\hline
\end{tabular}
\caption{Time resolution of the Be-Be counters candidates.}
\label{times_BeBe}
\end{table}


\section{Strip Efficiency of the BM@N Barrel Detector~(BD)}\label{sec.Eff_BD} 

\begin{figure}[t]
\begin{center}
\includegraphics[width=1.\textwidth]{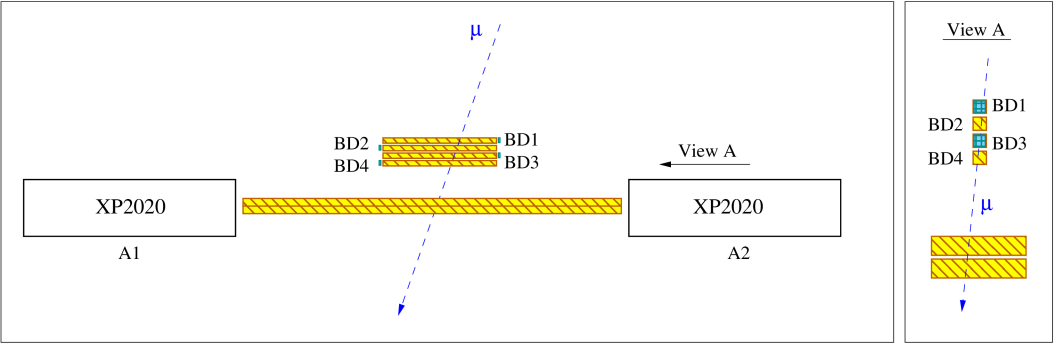}\\
\caption{Placement of the Barrel Detector strips for efficiency test.}
\label{Sfig8}
\end{center}
\end{figure}

The BD was fully assembled and used for the runs in 2017 and 2018 with beams of Carbon, Argon, and Krypton. Four spare strips were used in the current analysis. From these four strips, all of them wrapped in Al-mylar, three looked visually clean but the fourth one, with additional optical grease left from some previous tests, looked dirty. It was expected that light collection in this strip would be worse than in the other three; nevertheless, it was also included in the analysis.
The geometry of the setup of the strips and the trigger counter is shown in fig.~\ref{Sfig8}.

\begin{figure}[b]
\centering
\includegraphics[width=0.7\textwidth]{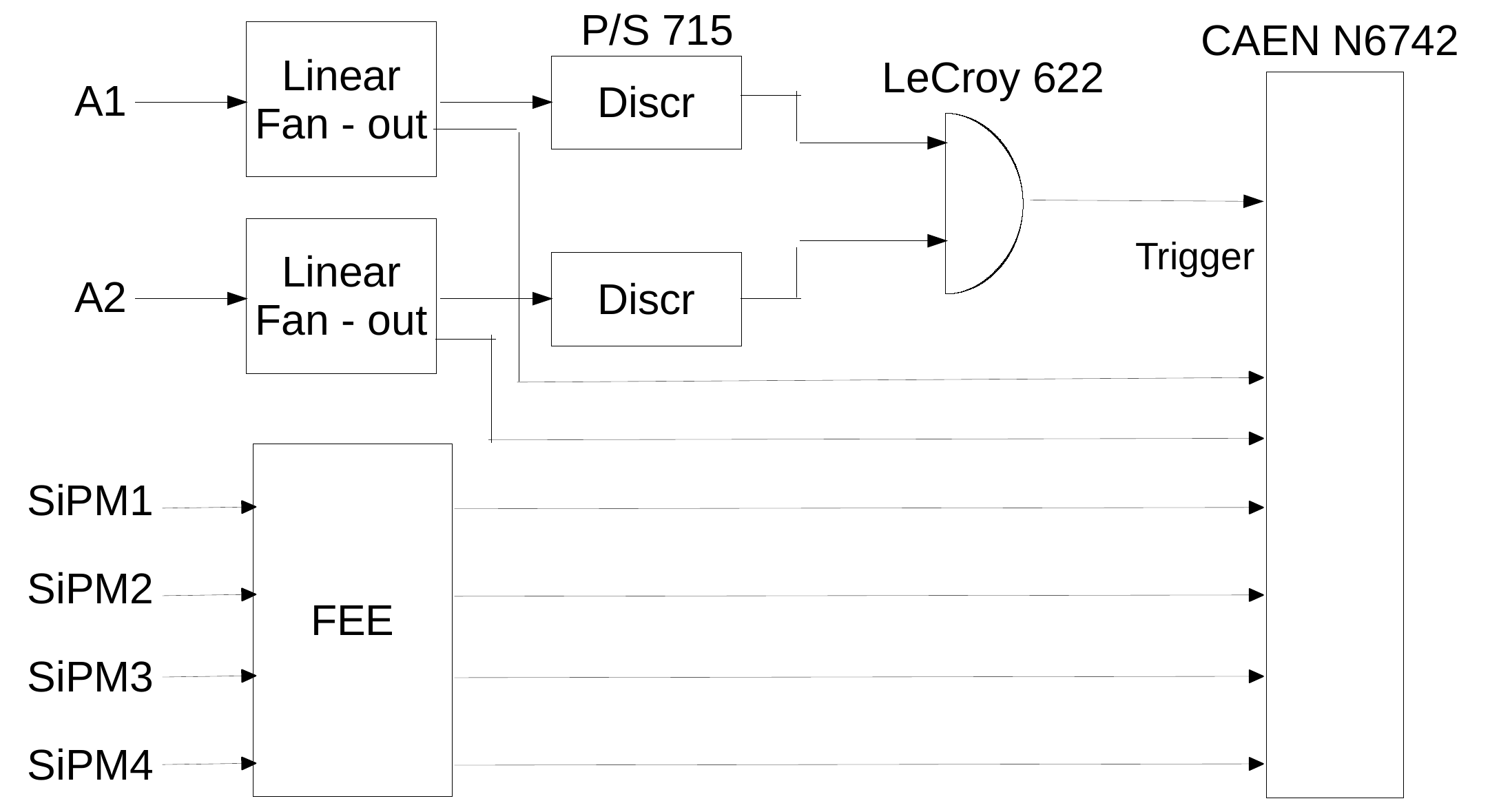}
\caption{Read-out electronics for the BD strips.}
\label{BDread}
\end{figure}

The BD strip consists of one SiPM Sensl Micro FC-60035-SMT coupled to a plastic scintillator BC418 wrapped in Al-mylar of $150\times7\times7~mm^3$. This SiPM has $6 \times 6~mm^2$ sensitive area and $35~\mu m$ microcell size~($18980$ microcells). The breakdown voltage is $24.5~V$, while the operating voltage is typically set above the breakdown voltage by $2-5~V$~(overvoltage). The Fast Output was used for the signal read-out. In this mode, the rise time of the output pulses is about $3~ns$. The overvoltage of the SiPMs was set to $1.75~V$, which corresponds to the voltage used in the BM@N run in March-April 2018. Similarly, the voltage for the two amplification stages in FEE was set for both to $5.000~V$, as in the experiment. The fig.~\ref{BDread} shows schematically the read-out electronics, where SiPM1, SiPM2, SiPM3, and SiPM4, correspond to BD1, BD2, BD3, and BD4, respectively. 

Tested counters were placed above a trigger counter, which provided the start signal for data read-out. Two $50\times5\times1~cm^3$ scintillator strips coupled to two PMTs~(Philips XP2020) were used for the trigger counter. The PMTs coincidence provided a trigger for a CAEN N6742 digitizer. Thus, two channels were used to sample negative pulses from the trigger PMTs and four channels to sample positive pulses from the SiPMs of the test detectors. The sampling frequency of the digitizer was set to the highest level~($5~GHz$, $1024$-bin-long waveform, $0.2~ns$ time bin).

\begin{figure}[t]
\begin{center}
\includegraphics[width=0.62\textwidth]{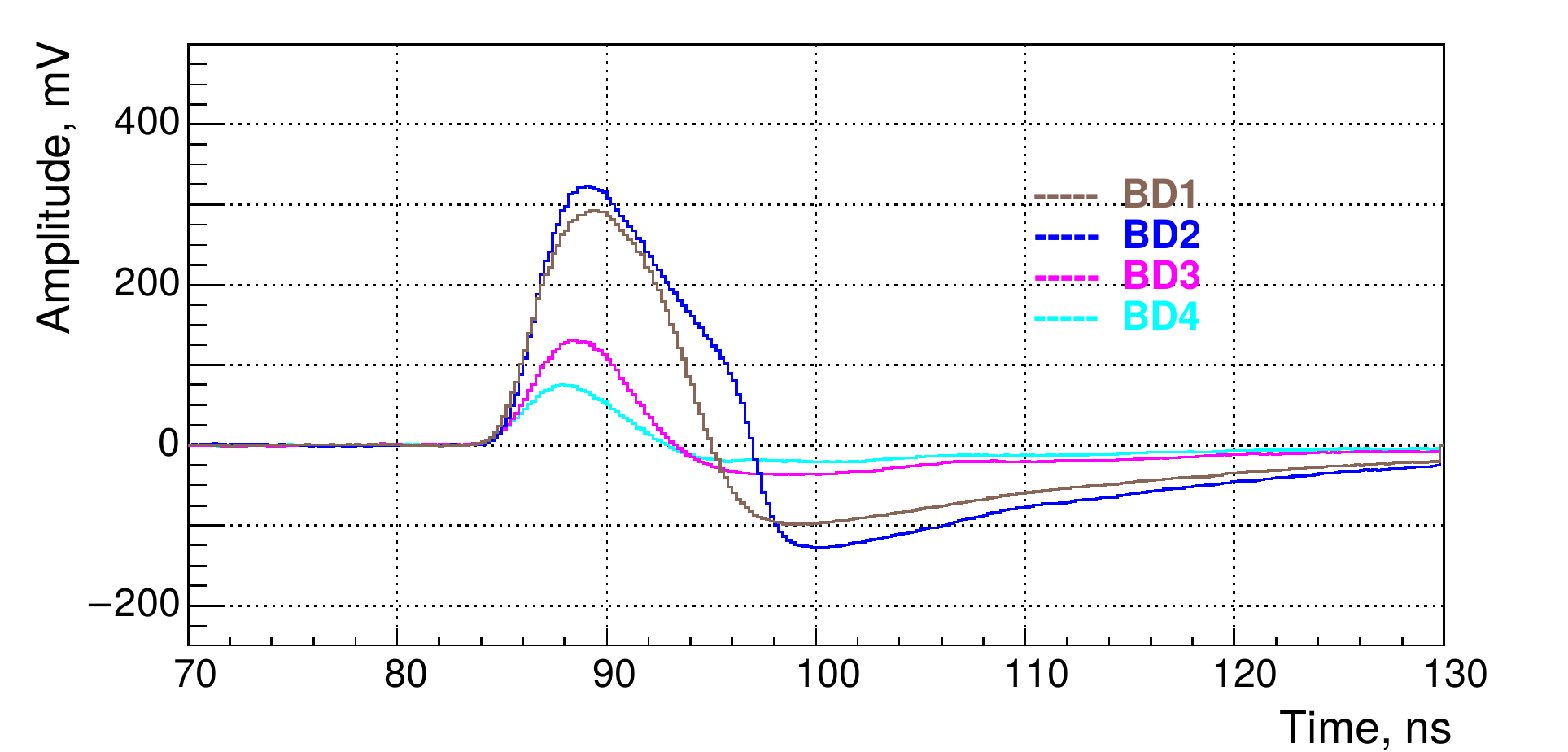} 
\includegraphics[width=0.62\textwidth]{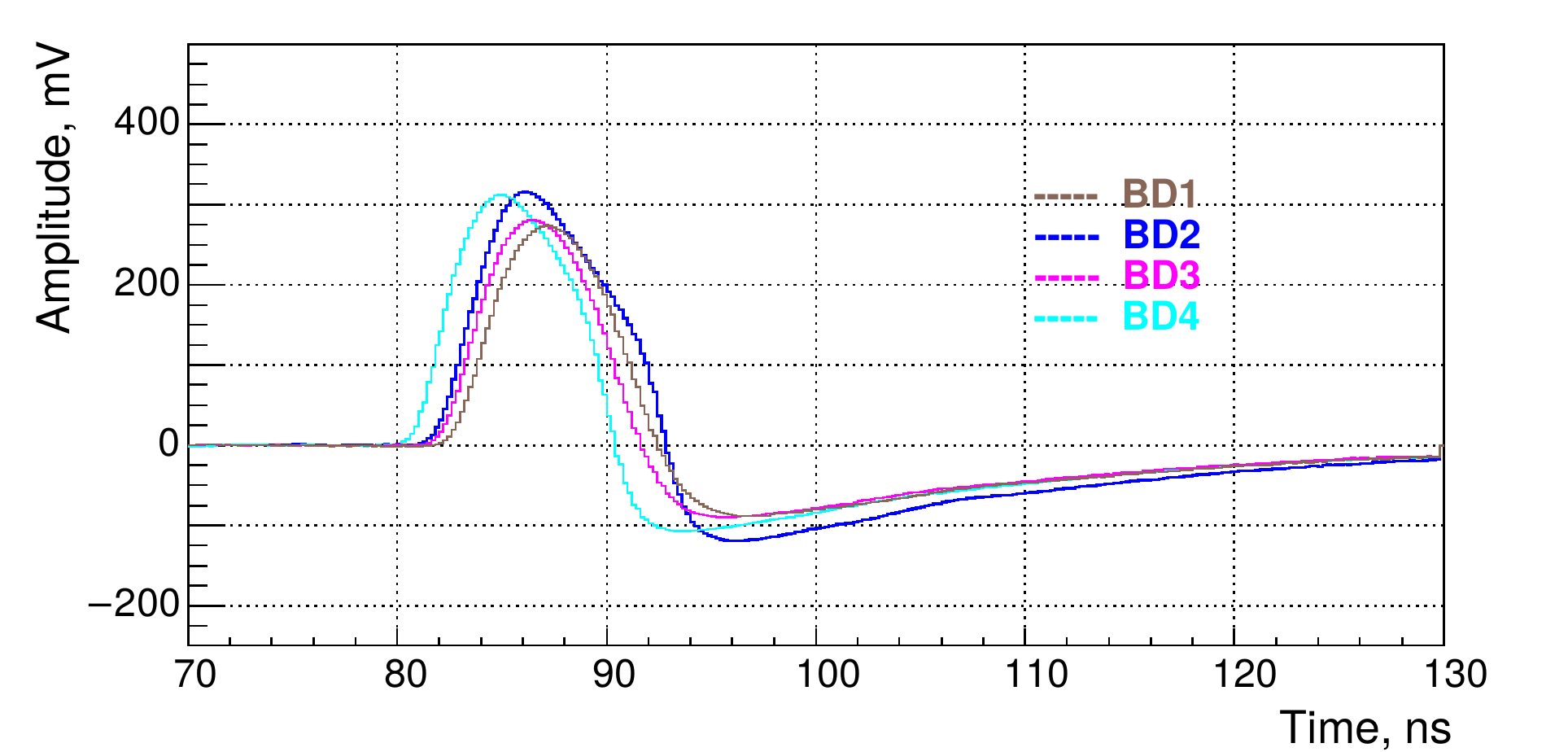}
\caption{Digitized pulses from BD Strips.}
\label{Sfig5}
\end{center}
\end{figure}

The selection criteria for the events correspond to signals in which at least three of the BD strips cross the threshold level at $25~mV$. Typical analog pulses from each of the BD strips are presented in fig.~\ref{Sfig5}.

\begin{figure}[H]
\begin{center}
\includegraphics[width=0.65\textwidth]{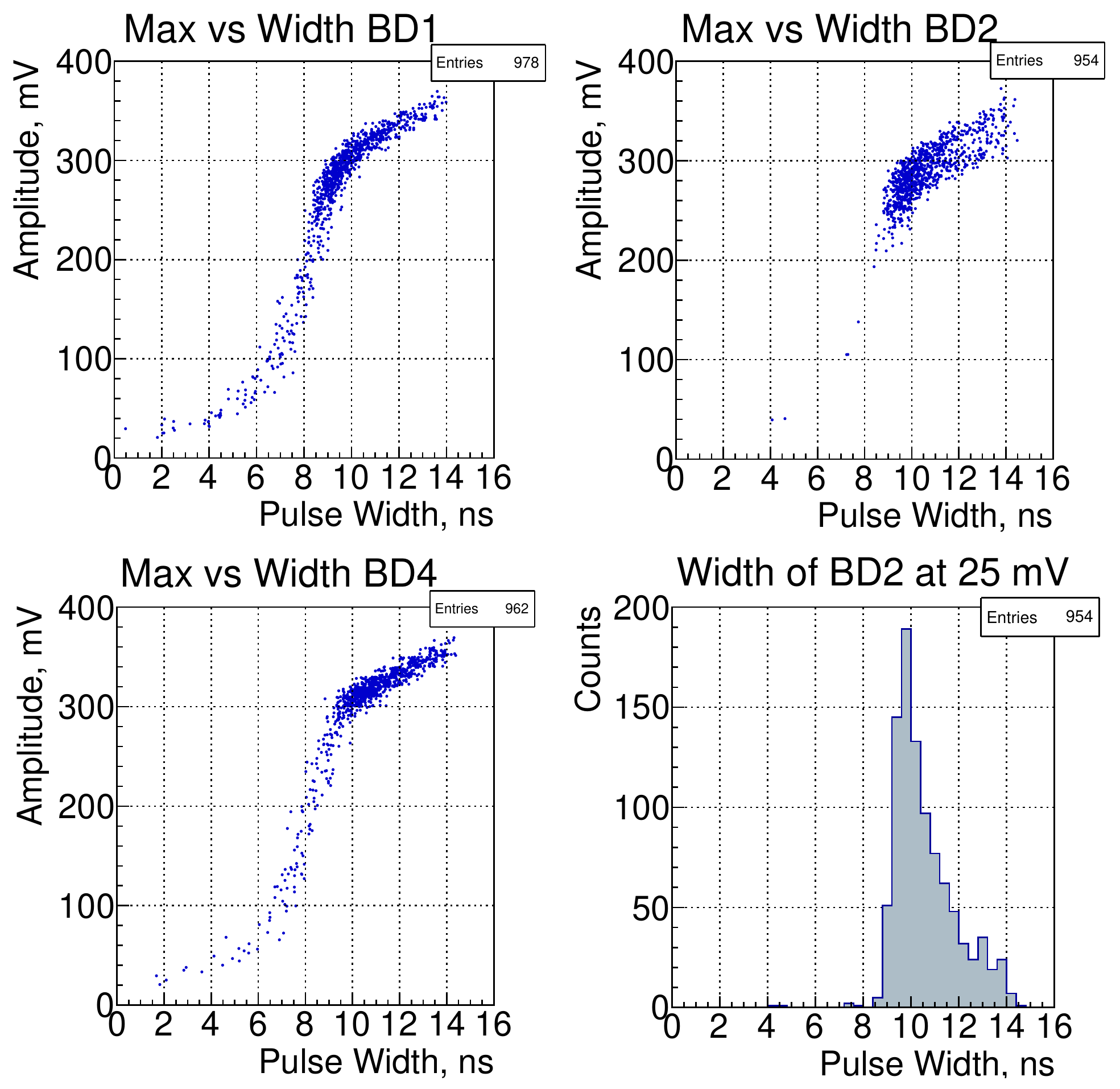}\\
\caption{BD Strips: pulse width versus amplitude.}
\label{BDWidth}
\end{center}
\end{figure}

The readout of the BD in the BM@N experiment is performed by using multi-hit time to digital converters (TDCs), which are based on the High-Performance Time to Digital Converter~(HPTDC) developed in CERN. 
The TDCs registered the pulse arrival time~(the time when the wavefront crosses a certain threshold) and its width~(the difference between times when the tail and the wavefront cross the threshold). This method of measuring pulse amplitude via its width is called "time over threshold"~(ToT) and, in principle, offers larger dynamic range~(i.e., large pulses can saturate maximum limit of waveform digitizer,  while width would still reflect the amplitude of the pulse). In order to simulate the ToT method in the current study, we used as a measure of pulse height not only the pulse amplitude but also its width above a threshold of $25~mV$ (see fig.~\ref{BDWidth}).

\begin{figure}[H]
\begin{center}
\includegraphics[width=0.65\textwidth]{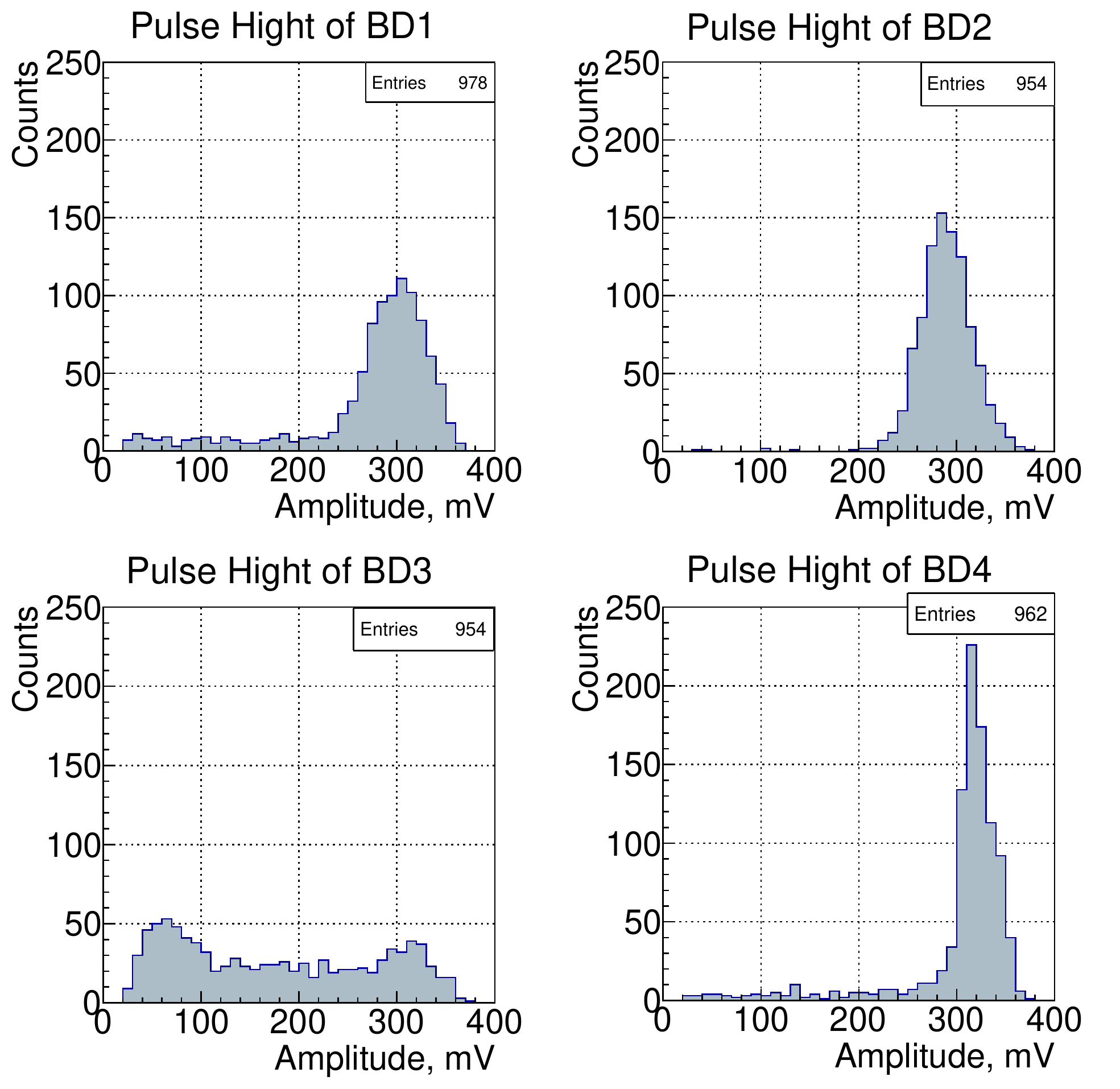}\\
\caption{Amplitude distributions of the signals from BD strips. See text for details.}
\label{Sfig9}
\end{center}
\end{figure}

Amplitude distributions from four tested BD strips are shown in fig.~\ref{Sfig9}. They were obtained with the following conditions. Pulses from every strip had to exceed $25~mV$. 
In addition, the distributions for BD2 and BD3 were filled when amplitudes in BD1 and BD4 exceeded $100~mV$, the distribution for BD1 was filled when amplitudes in BD2 and BD4 exceeded $100~mV$, and the distribution for BD4 was filled when amplitudes in BD1 and BD2 exceeded $100~mV$. As expected, the strip BD3, dirty with optical grease, showed poor amplitude resolution due to light loss. The other three strips presented better resolution. 
The amplitude distribution for the BD2 is the most representative due to the geometry of the setup and event selection (BD2 is placed between the BD1 and BD4).
With this result, one can state that detection efficiency for minimum ionizing particles if they entirely cross the BD strips should be close to $100\%$.

\newpage
\section{Time resolution of the BD counters}\label{sec.Time_BD}
\subsection{Experimental setup}

\begin{figure}[b]
\centering
\includegraphics[width=1.\textwidth]{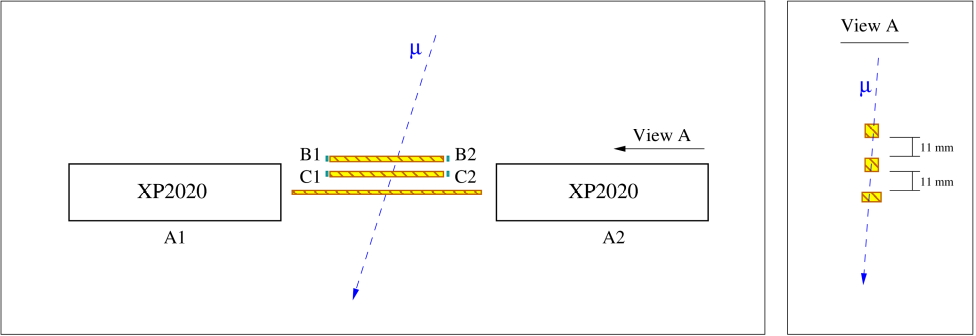}
\caption{Placement of counters in the evaluation of BD counters.}
\label{BDsetup}
\end{figure}

The measurements of the time resolution of the BD counters were performed using two of the three clean BD strips selected according to the efficiency test performed. The trigger counter consisted of a single scintillator bar $240\times10\times5~mm^3$~(also BC418) and two PMTs. The experimental setup is shown in fig.~\ref{BDsetup} with tested counters placed above the trigger counter. The light produced in each BD strip is detected by two SiPM Sensl Micro FC-60035-SMT coupled on the edges. Two sets of measurements for the SiPMs coupled to scintillators were done: with and without optical grease~(Rhodorsil). In both cases, the bias voltage of the SiPMs was set to $26.5~V$. The read-out electronics is the same as used in the efficiency test already shown in fig.~\ref{BDread}, In this case, SiPM1, SiPM2, SiPM3, and SiPM4 are now B1, B2, C1, and C2, respectively.

Since the counters were placed in a large black box, we also performed a temperature test. The data acquisition lasted several hours during daytime, and it was necessary to verify the heating influence in the measurements. The SiPM breakdown voltage during several-hours-long data taking, together with its front-end electronics and the voltage dividers of the PMTs, heated the black box containing all these components. Therefore, it was necessary to measure the temperature inside and outside the box to compare it, before and after data accumulation. No difference in the temperature inside and outside of the box at the end of the data taking period was detected~(with $1^o~C$ precision). 

\begin{figure}[h]
\centering
\includegraphics[width=0.49\textwidth]{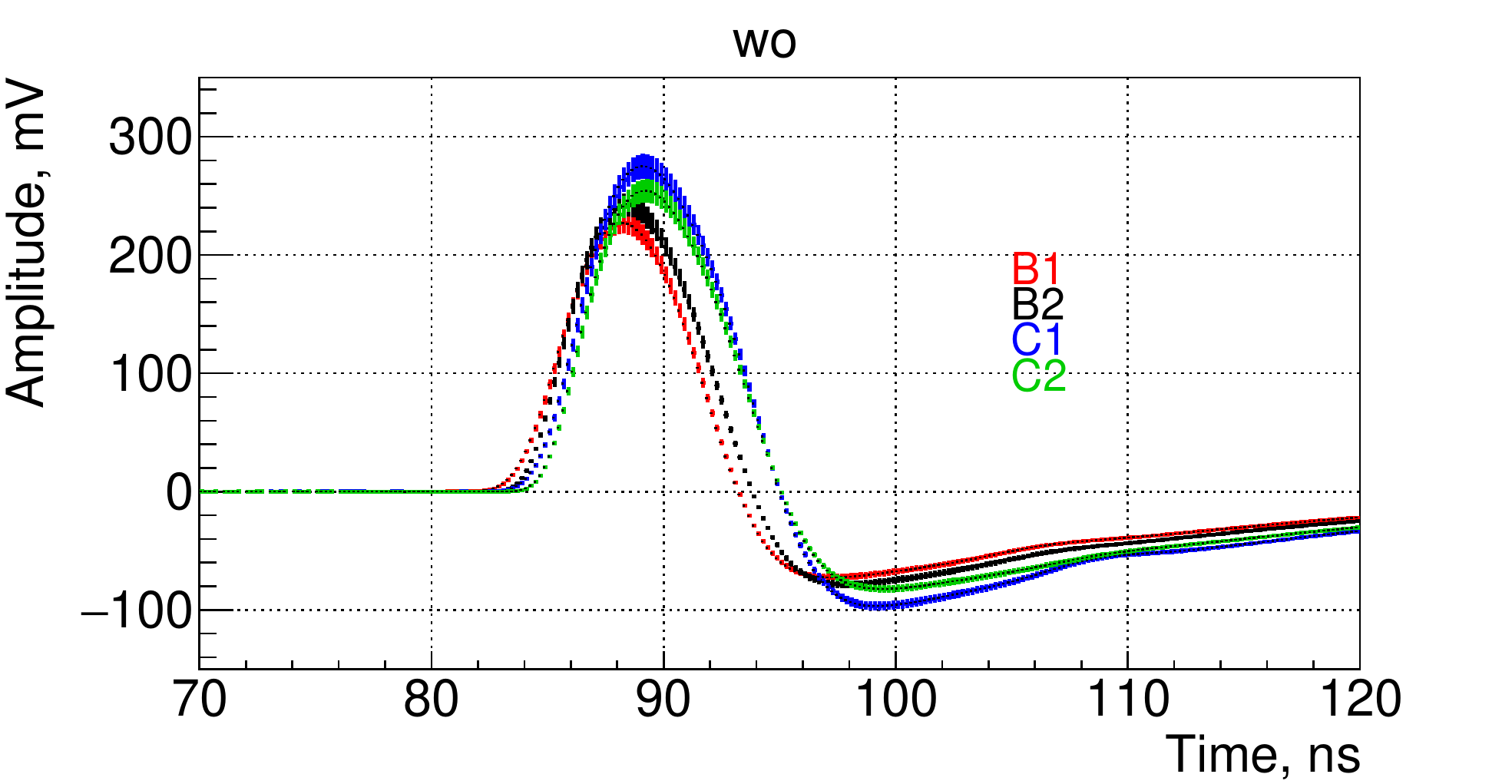}
\includegraphics[width=0.49\textwidth]{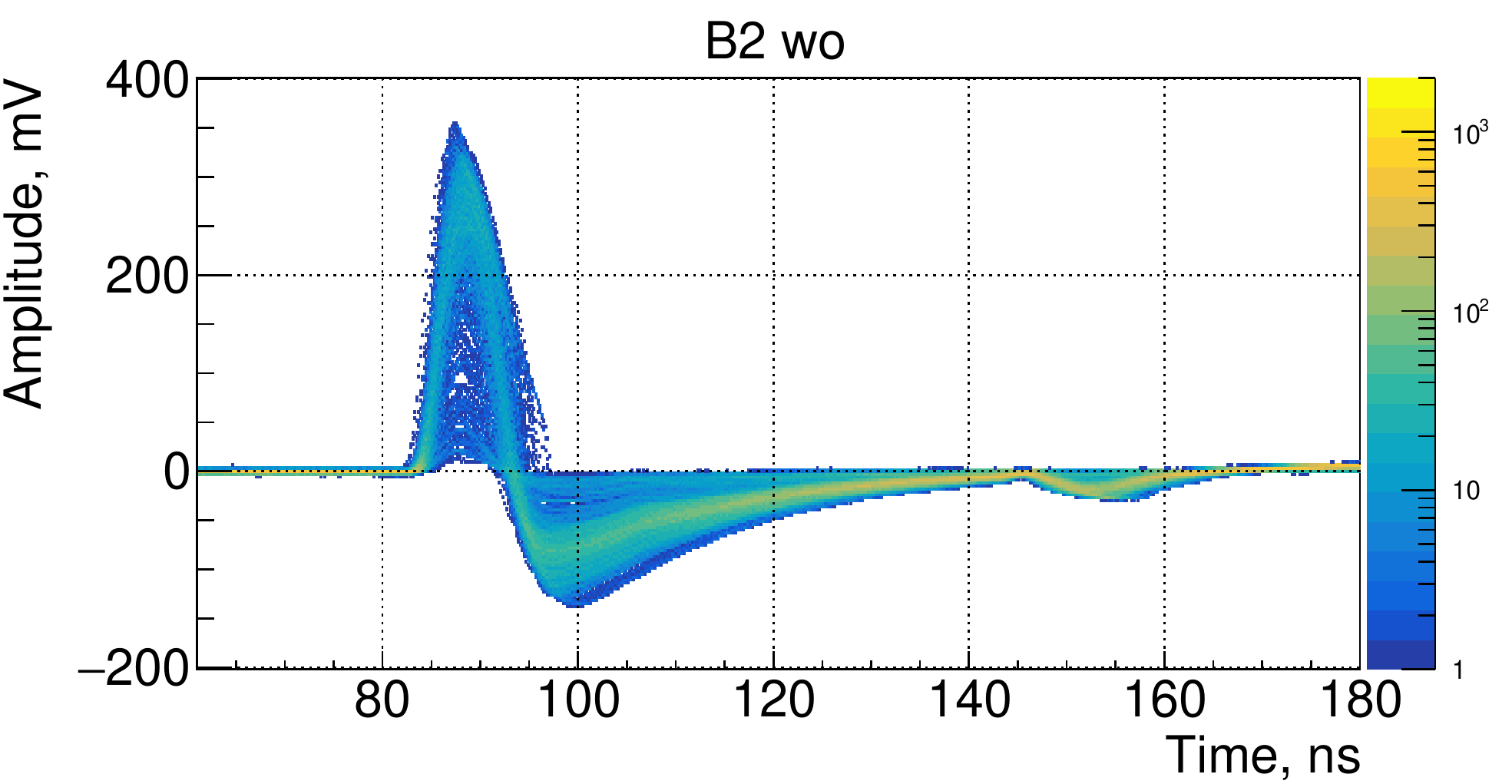}
\includegraphics[width=0.49\textwidth]{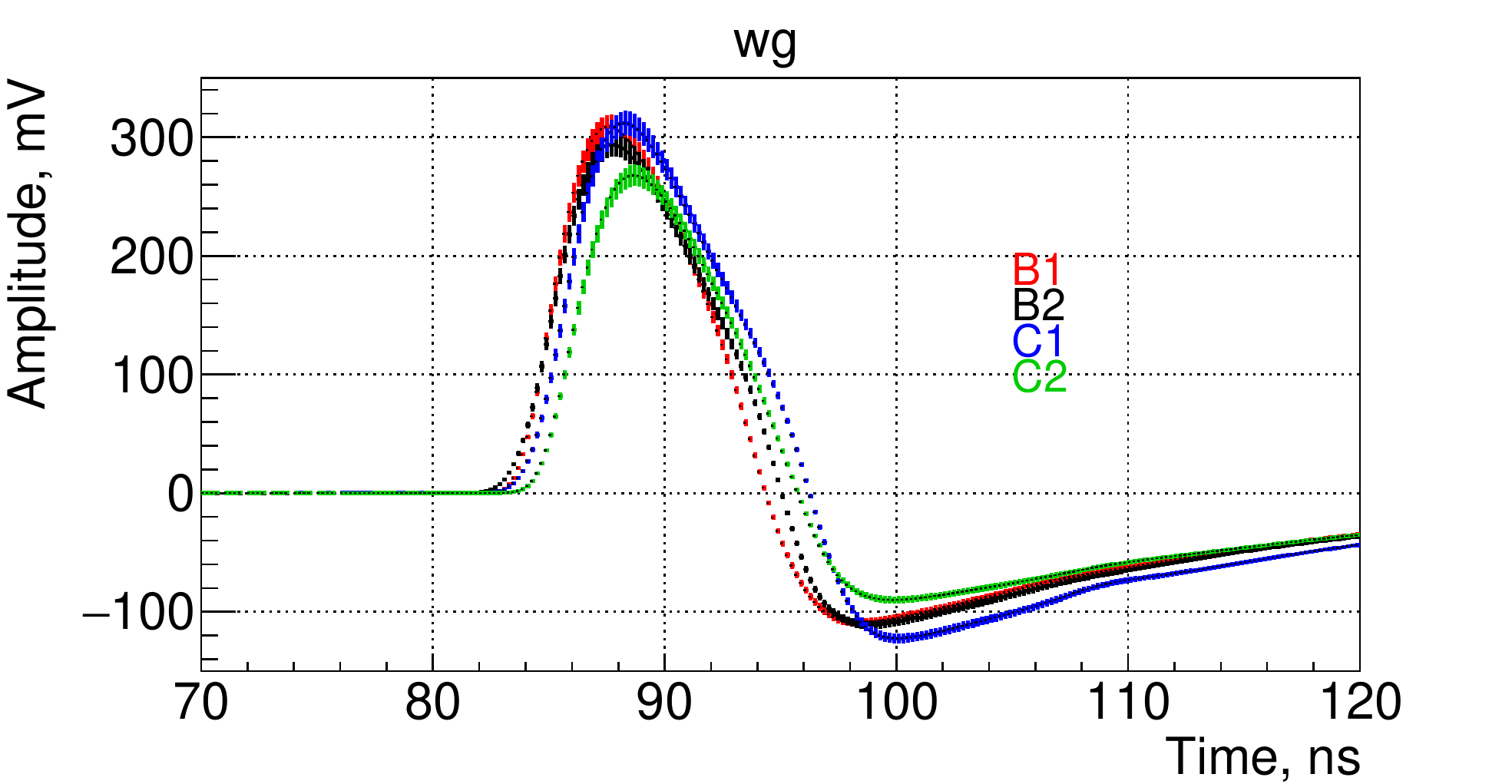}
\includegraphics[width=0.49\textwidth]{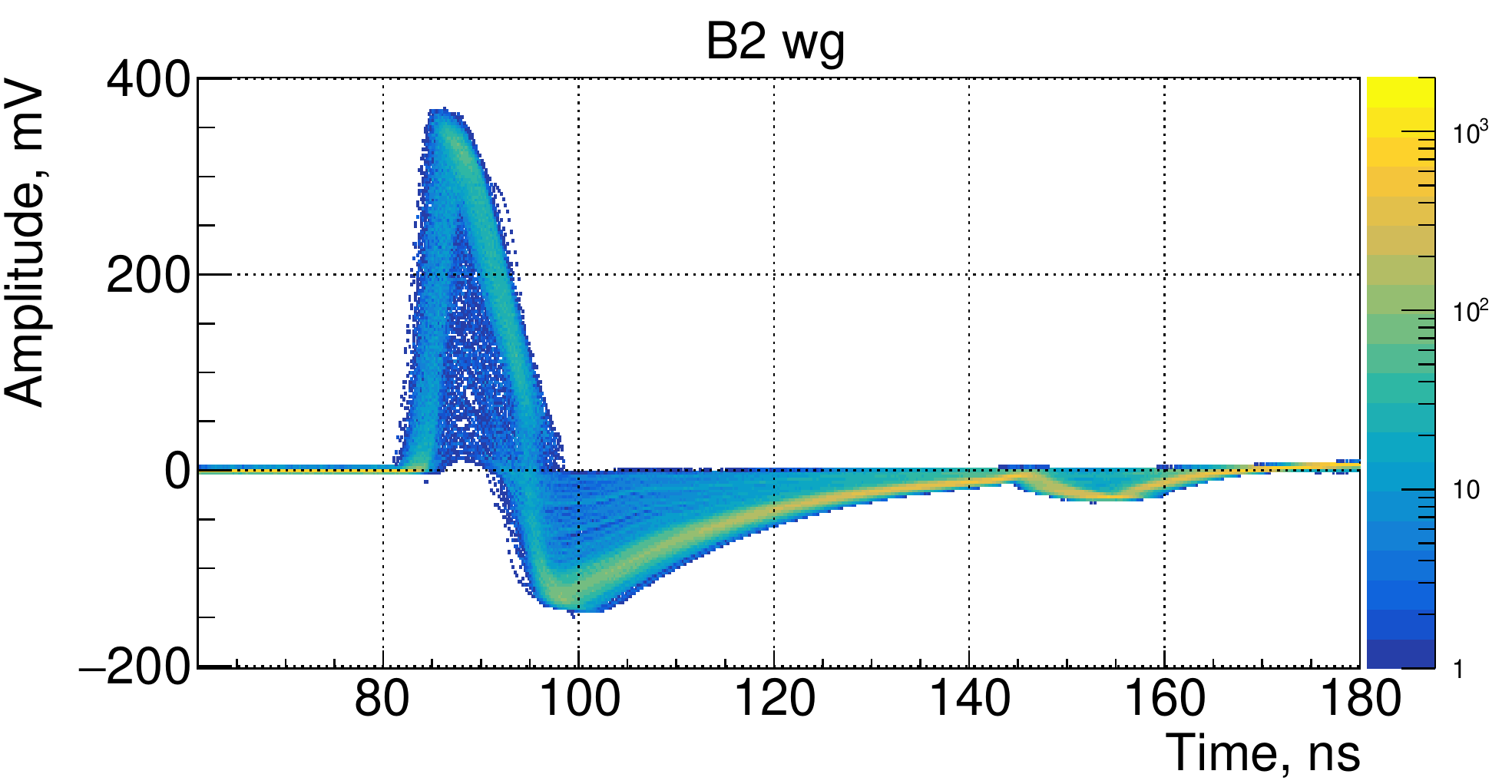}
\caption{Mean signal for the BD counters with~(bottom-left) and without~(top-left) optical grease.
Persistence of the B2, with~(bottom-right) and without~(top-right) optical grease.}
\label{Mean_BD}
\end{figure}

Fig.~\ref{Mean_BD}~(left) shows the mean signals from the B1, B2, C1 and C2 with~(bottom) and without~(top) optical grease.
The pulse amplitude increased more than 10\% when optical grease is applied. 
No noise or after pulses were observed in the persistence plots~(right).
Table~\ref{T_MeanS_BD} shows the properties obtained from the mean signals, with special attention to the differences  in the pulse amplitudes due to the optical grease effect.
 
\begin{figure}[H]
\includegraphics[width=0.33\textwidth]{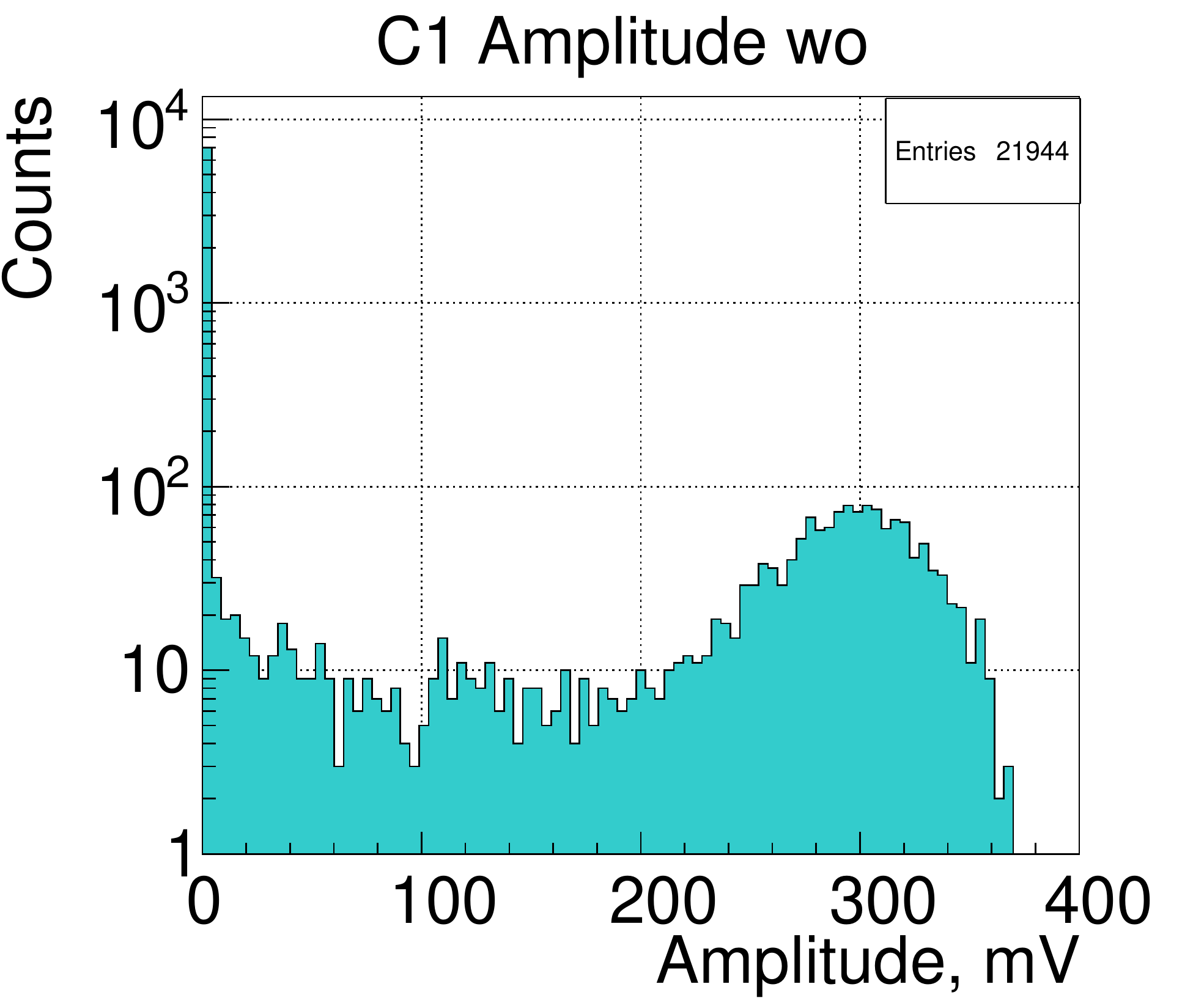}
\includegraphics[width=0.33\textwidth]{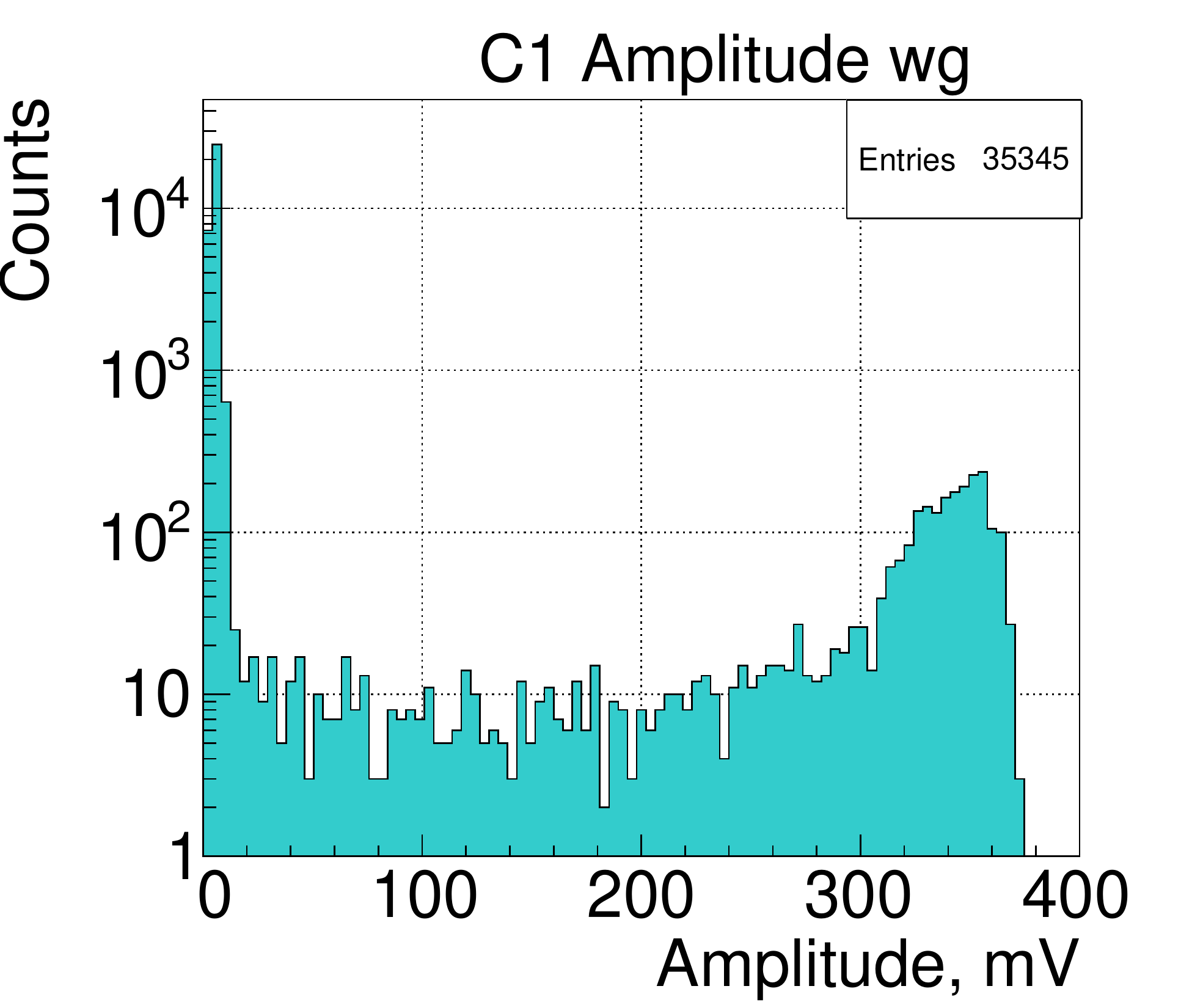}
\includegraphics[width=0.33\textwidth]{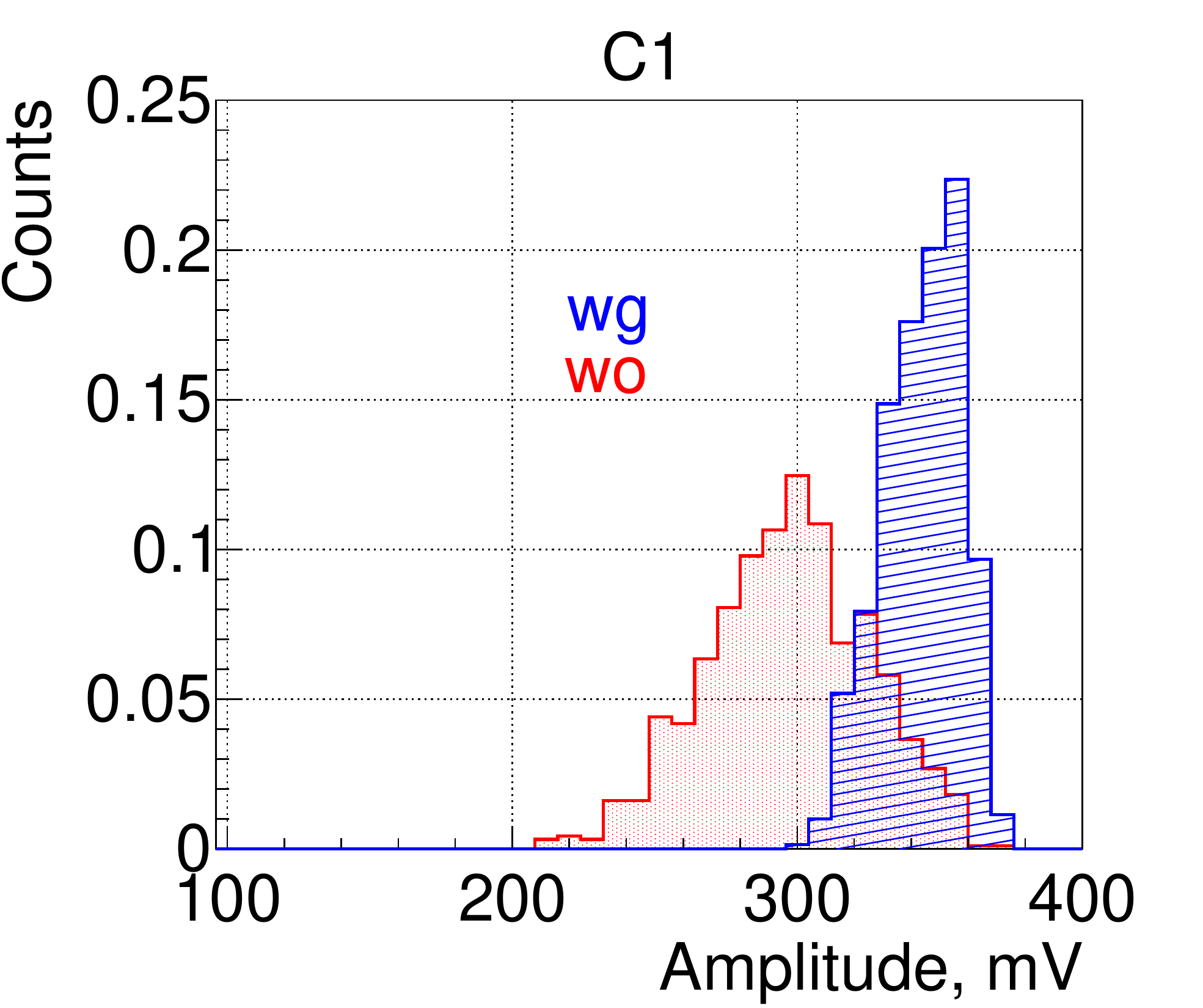}\\
\caption{Amplitude distributions of BD counters with~(center) and without~(left) optical grease. Coincidence distributions~(right) with~(blue) and without~(red) optical grease over threshold.}
\label{Amp_BDtr}
\end{figure}

Fig. \ref{Amp_BDtr} shows the amplitude distributions for C1 with and without optical grease, similar distributions were obtained for the other SiPMs. 
The distinction between noise and signals is clear due to high efficiency of light collection.
The threshold was set at $300~mV$ and $200~mV$ for the case with and without optical grease, respectively. 
Events in which pulses from all the SiPMs exceeded the threshold value were selected for the time
resolution analysis.
The amplitudes after the selection criteria are shown in fig. \ref{Amp_BDtr}~(right), the distributions were normalized with the total number of selected events~(integral).

\begin{table}[htbp]
\centering
\begin{tabular}{|c|c|c|}
\hline
Properties & wo & wg \\\hline\hline
Amplitude, mV  & $248\pm5$&$294\pm5$ \\\hline
Rise Time, ns  & $3.1\pm0.2$&$2.8\pm0.2$ \\\hline
Peak width, ns & $6.3\pm0.4$&$7.2\pm0.3$ \\\hline
\end{tabular}
\caption{Signal properties of the BD counters, considering the average value from the mean signal in each SiPM.}
\label{T_MeanS_BD}
\end{table}

\newpage
\subsection{Time resolution measurement for the BD counters}

\begin{figure}[b]
\centering
\includegraphics[width=.85\textwidth]{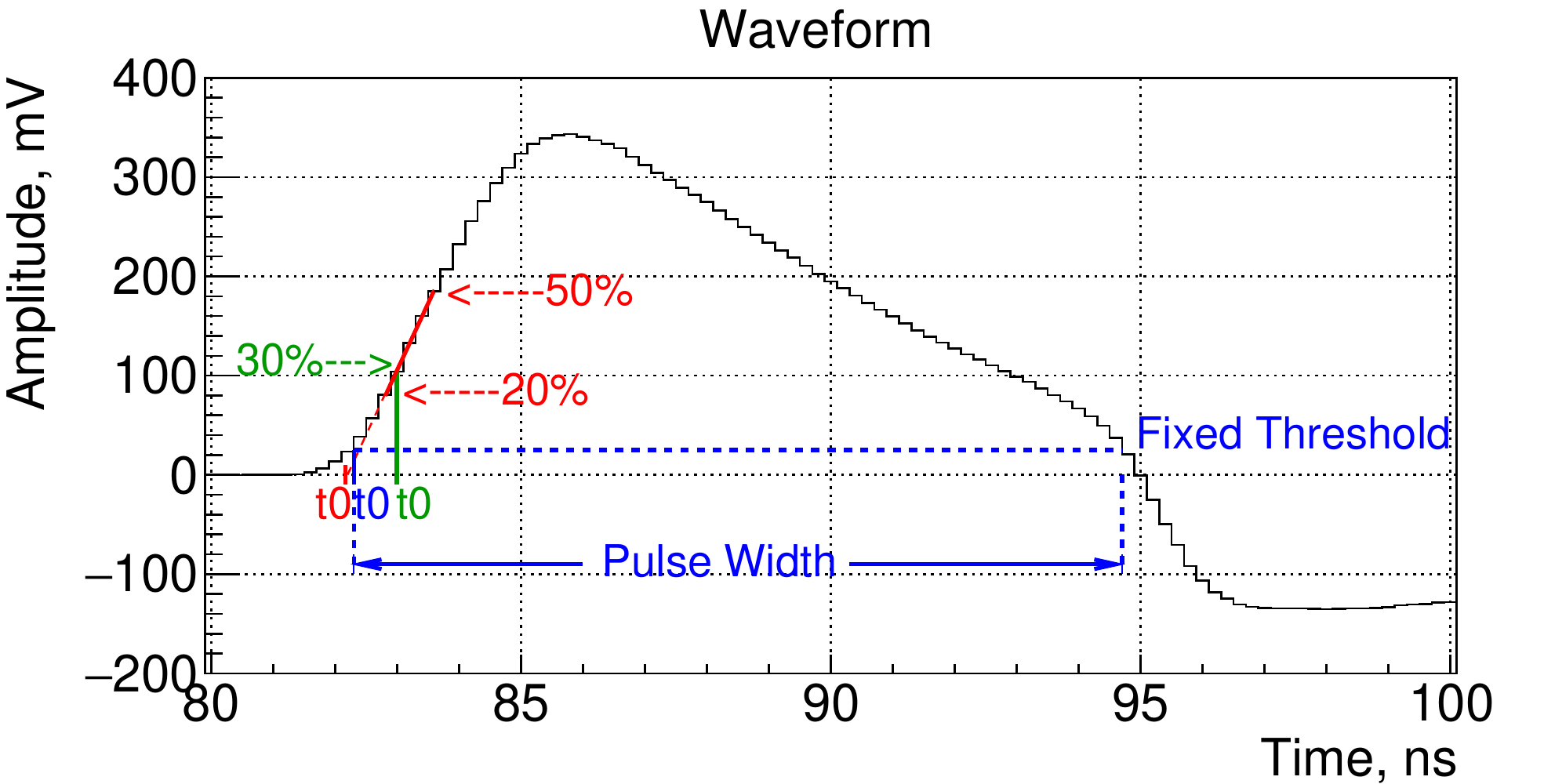}
\caption{Determination of the signal arrival time by CFD~(red and green) and FT~(blue). See text for details.}\label{methBD}
\end{figure}

We evaluated two basic approaches to calculate the arrival time~(t0) of the signals: CFD and fixed threshold~(FT). These methods are usually implemented on the hardware level in the discriminators. If the pulse shape does not depend on the amplitude, the FT will give the initial spread in the signal arrival time, while CFD does not present this shift. Fig.~\ref{methBD} shows the methods M1~(red) and M4~(green). The description of these methods was already given in section~\ref{sec.TRBEBE}. In blue~(M5), a threshold level is defined~(25~mV) to evaluate the FT approach, where t0 is the time
when the signal cross this threshold. 
We evaluate four different thresholds levels: $5$, $10$, $16$, and $25~mV$ to determine t0 and the signal width by the ToT method. Note that t0 varies depending on the amplitude, so it had to be adjusted with slewing correction (or time walk correction).  
In the slewing correction we used the pulse width as a measure of the pulse height.

\begin{figure}[t]
\includegraphics[width=0.49\textwidth]{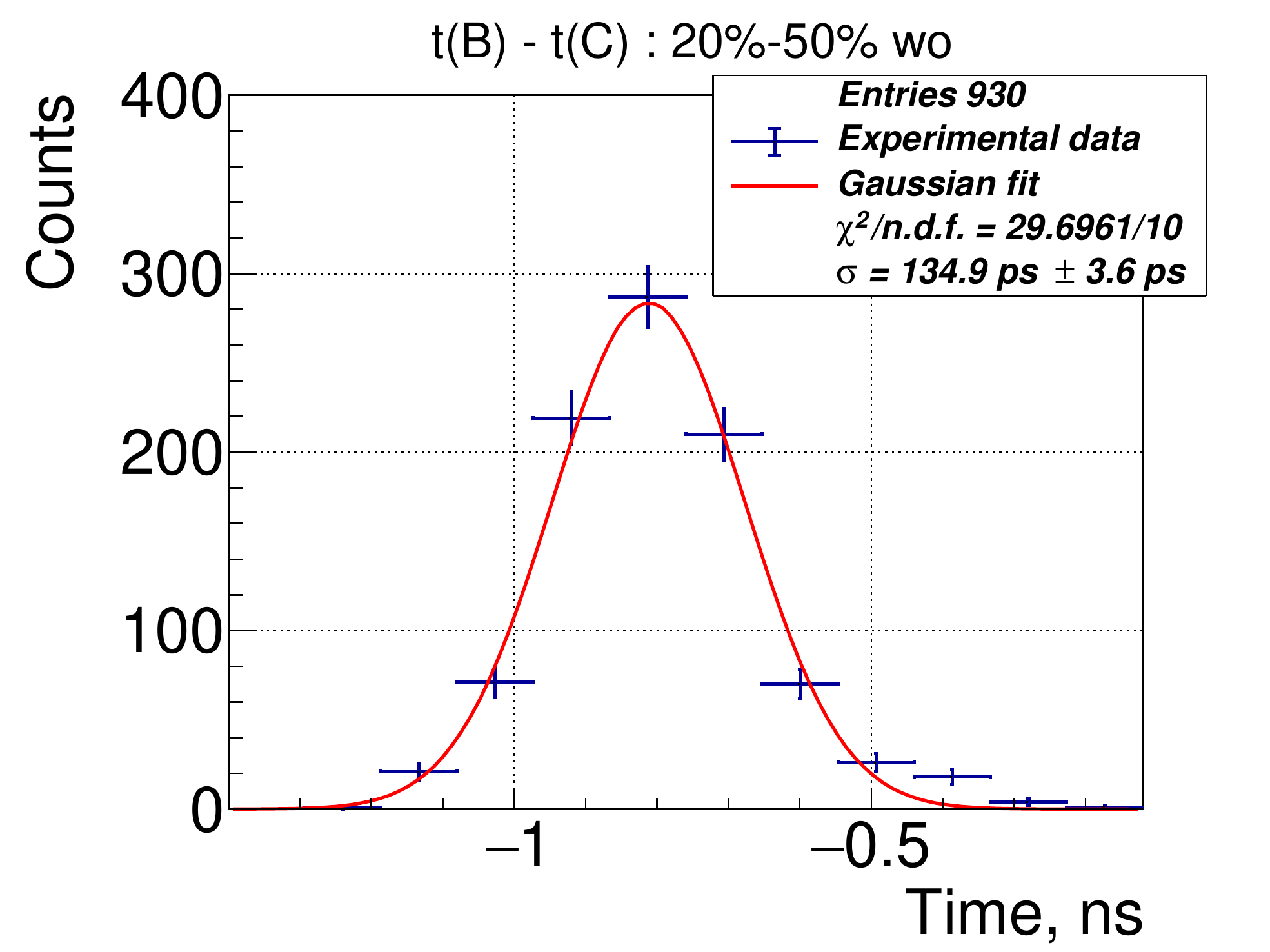}
\includegraphics[width=0.49\textwidth]{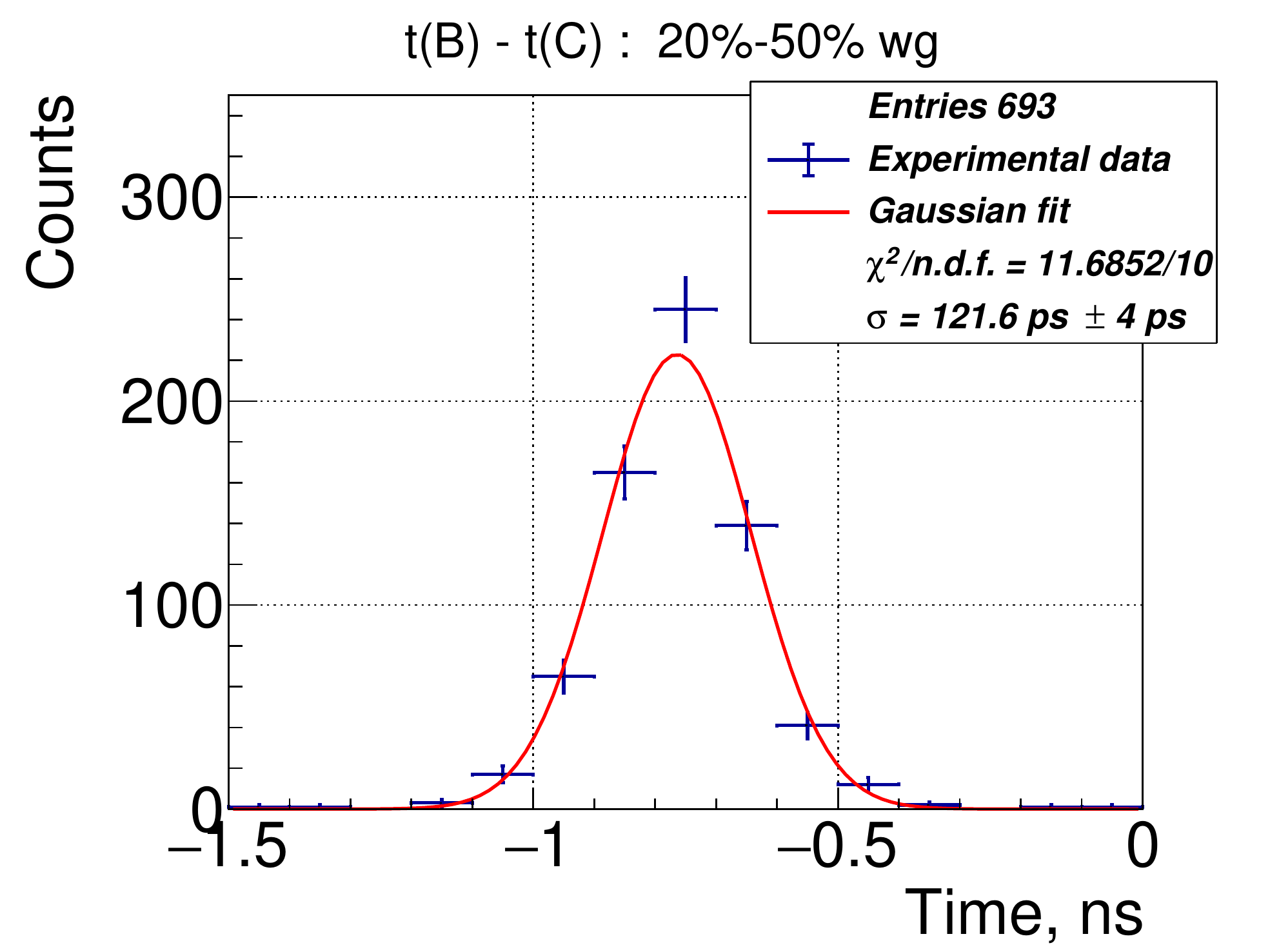}\\
\caption{Time difference distributions of BD counters with~(right) and without~(left) optical grease.}
\label{timesD_BD}
\end{figure}

The impact times in the detectors A, B, and C were calculated as $t(A) = \frac{1}{2} (t0_{A1}+t0_{A2})$, 
$t(B) = \frac{1}{2} (t0_{B1}+t0_{B2})$, and $t(C) =\frac{1}{2}  (t0_{C1}+t0_{C2})$, respectively. 
Fig.~\ref{timesD_BD} shows the distribution of the difference $t(B) - t(C)$ obtained by the method M1, 
where the standard deviation corresponds to $\sigma_{BC}$~(the time resolution between B and C). Similar results were obtained with M2, M3, and M4. The time resolution between pairs of counters A and B~/~A and C  was also determined under M1~($242\pm8~ps~/~246\pm8~ps$ with and $250\pm7~ps~/~248\pm7~ps$ without optical grease).

As was mentioned in section~\ref{sec.Eff_BD}, the ToT method gives the times when front and tail of the signal cross the fixed threshold.
The difference between these times defines the width of the pulse. Then, to model the ToT method, it was decided that the signal width define the selection criteria for M5, considering only the events when all the SiPMs signals had a width greater than zero.

\begin{figure}[tb]
\centering
BD counters-B2 wg\\
\includegraphics[width=1.0\textwidth]{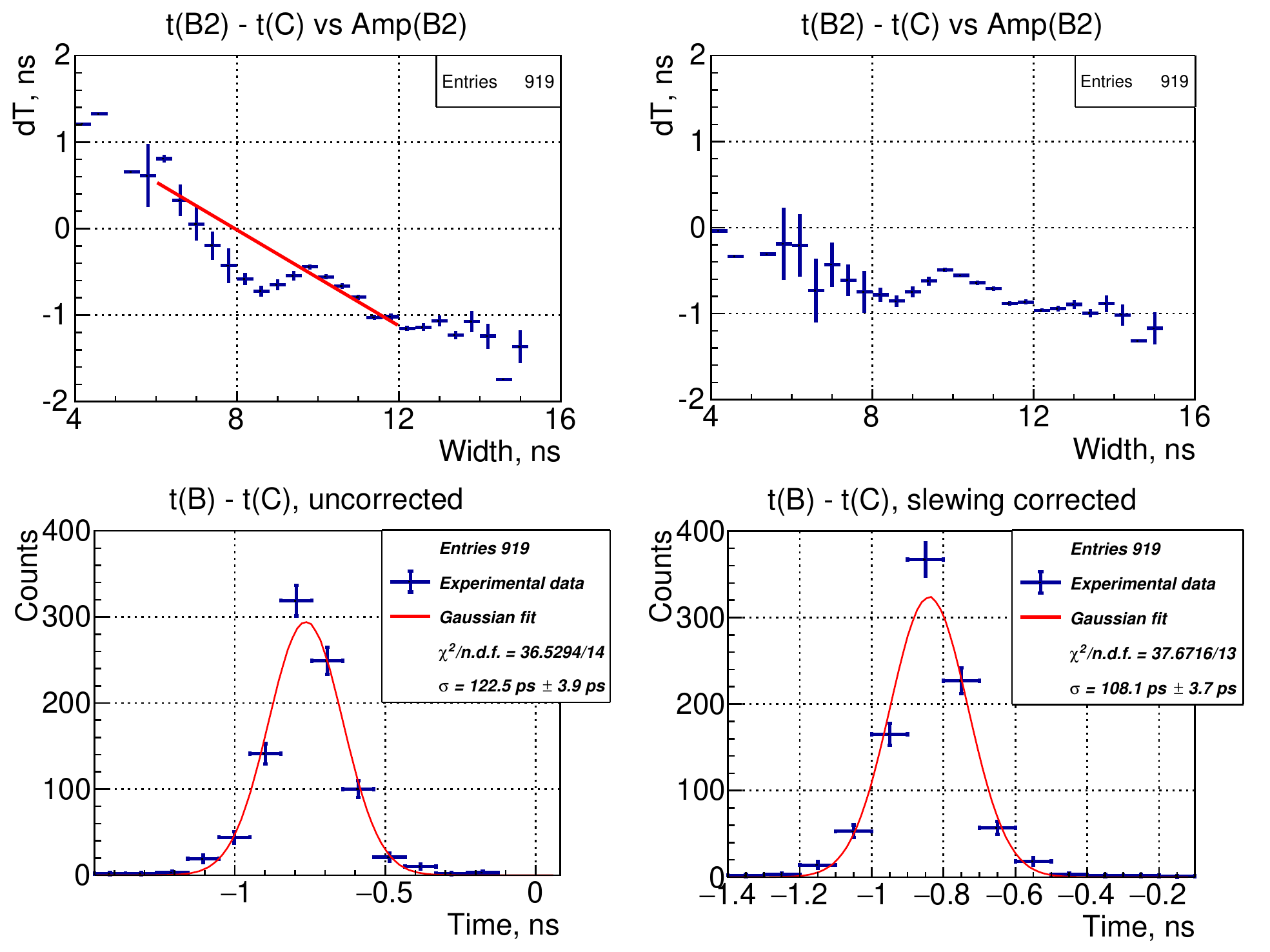}
\caption{The distributions for the FT method before~(top-left) and after~(top, right) the slewing correction for B2 with optical grease~(wg). The Gaussian fits for the time differences before~(bottom-left) and after~(bottom, right) the correction.}
\label{SC}
\end{figure}

As we can see in fig.~\ref{SC}~(top-left) the slewing correction was done for the B2~(wg) by plotting the dependence of the time difference~(dT) between, t0(B2) and (t0(C1)+t0(C2))/2 as a function of B2 width and then fitting this dependence with a straight line. The time spread of the uncorrected dT is more than 2~ns. After the subtraction of this fit function from raw t0(B2), we obtained the corrected t0(B2)~(top-right). Similarly, B1, C1, and C2 were slewing corrected. The slopes of the fit function for the four channels are close to each other.
After the correction, the distribution did not show a completely flat pattern, so, there is still a width dependence left. However, if one tries to make the pattern flatter, increasing the correction factor, one gets flatter "dT vs. Width" but worse time resolution. This may be due to the hit position dependence, which affect both, the pulse amplitude and light propagation time.
The time resolution obtained for the corrected and uncorrected t0 is shown in fig.~\ref{SC} at the bottom-right and bottom-left, respectively.

Table~\ref{times_BD} presents the time resolution obtained for a single BD counter by the methods evaluated under the assumption $\sigma_B = \sigma_C = \frac{\sigma_{BC}}{\sqrt{2}}$. 
The results indicate that after slewing correction FT method provides better time resolution that the CFD method. 
However, the slewing correction is normally performed offline and difficult to implement online.
It is a disadvantage compared to the CFD method, which can be performed online. 
As was indicated in section~\ref{sec.TRBEBE},  
when optical grease is applied, the light collection improves, which leads to a better time resolution.

\begin{table}[H]
\centering
\begin{tabular}{!{\VRule[2pt]}c|c|c|!{\VRule[2pt]}c|c|c!{\VRule[2pt]}}
\hline
{Method} & \multicolumn{2}{c|!{\VRule[2pt]}}{Time resolution~(ps)} & {Method 5} & \multicolumn{2}{|c!{\VRule[2pt]}}{Time resolution~(ps)} \\ \cline{2-3} \cline{5-6}
 CFD~(\%) & wo & wg & FT~(mV) & wo U~/~C & wg U~/~C \\
\hline \hline
M1~(20-50) & 
$95\pm3$ & $86\pm3$

& $25$ &
$144\pm6$ / $82\pm3$ & $92\pm3$ / $79\pm3$
	
\\\hline
M2~(10) & 
$93\pm3$ & $87\pm3$
 
& $16$ &
$127\pm5$ / $84\pm3$ & $93\pm4$ / $82\pm3$

\\\hline
M3~(20) & 
$104\pm4$ & $85\pm3$

& $10$ &
$92\pm4$ / $75\pm2$ & $88\pm3$ / $76\pm3$

\\\hline
M4~(30) & 
$110\pm4$ &	$88\pm4$

& $5$ &
$95\pm3$ / $74\pm2$ &	$87\pm3$ / $76\pm3$
\\\hline
\end{tabular}
\caption{Time resolution of the BD strips, where 
U and C corresponds to uncorrected and slewing corrected time resolution.}
\label{times_BD}
\end{table}

\begin{figure}[H]
\includegraphics[width=0.49\textwidth]{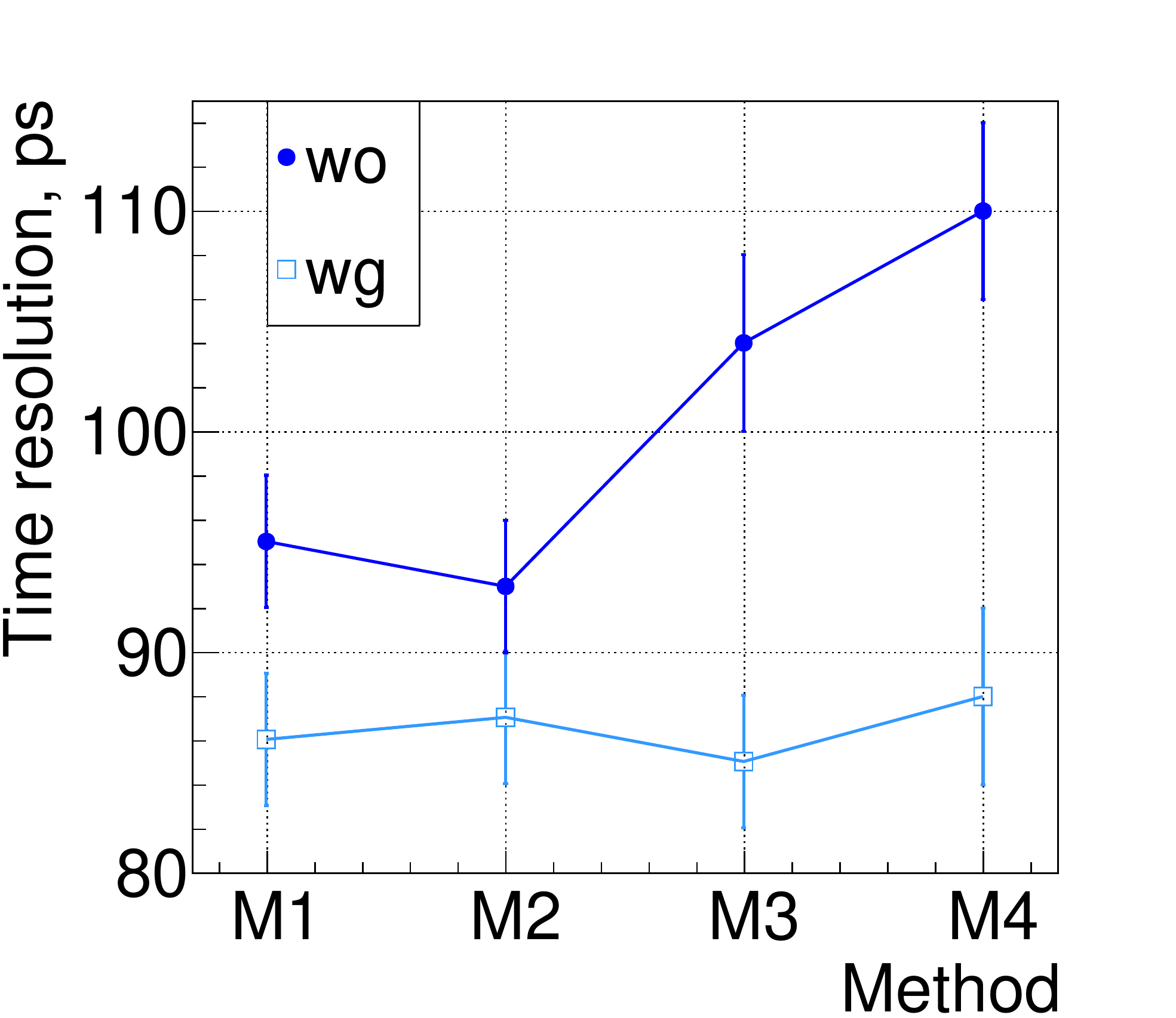}
\includegraphics[width=0.49\textwidth]{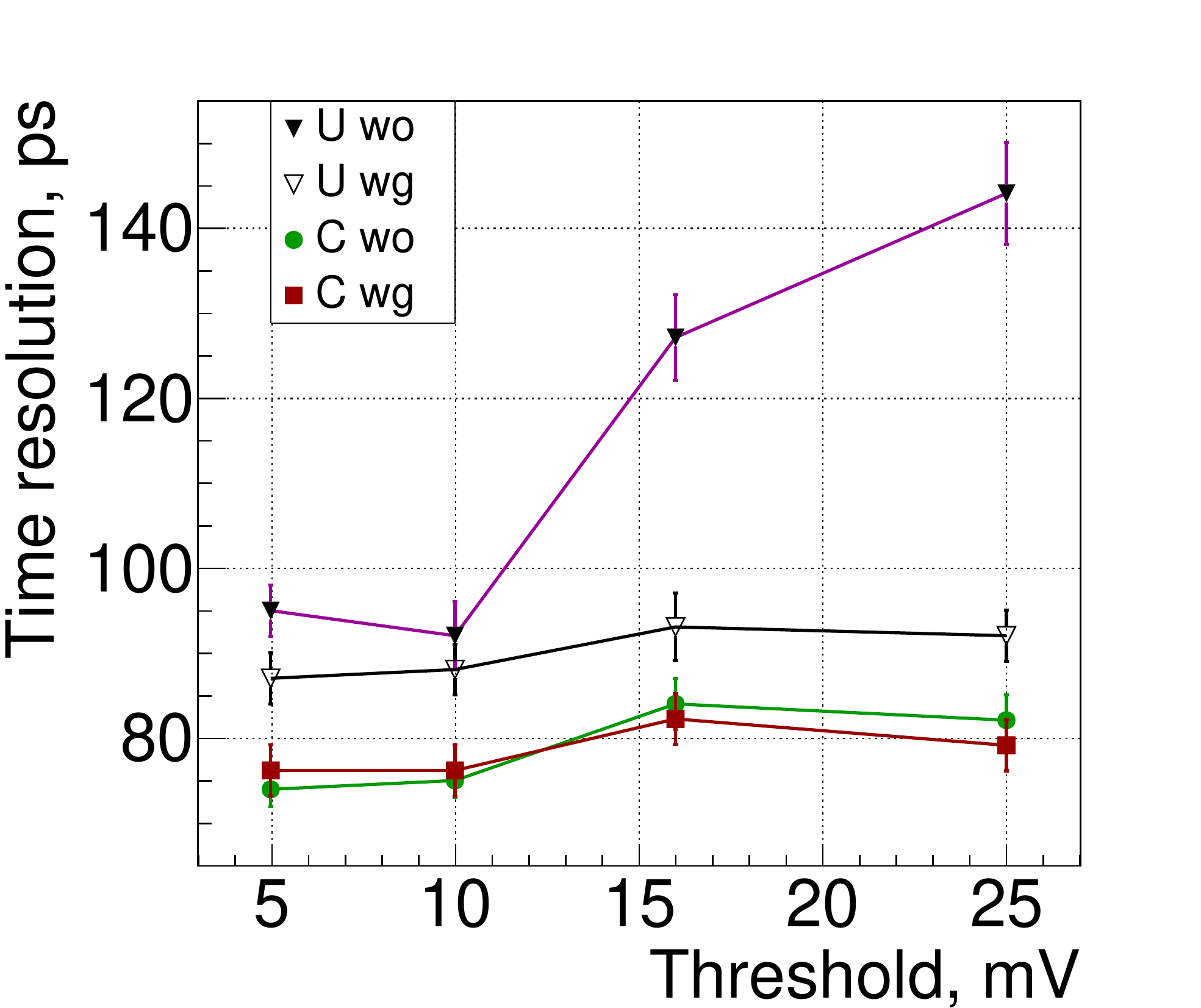}
\caption{Time resolution of each method evaluated (CFD~left plot, FT~right plot)
for the BD counter with and without optical grease.
For FT, the uncorrected~(U-black triangles) 
and slewing corrected~(C-red~wg and C-green~wo) time resolution 
for the different fixed thresholds.
}
\label{TR_BD}
\end{figure}

Figures~\ref{TR_Abs} and \ref{TR_BD} show the time resolution obtained with different methods evaluated in this analysis.
Compared to the Be-Be counter's time resolution values, which show a significant dependence on the method used to determine the arrival time of the signals, the resolution of the BD counters is less sensitive to the analysis method.
This difference can be attributed to shorter and statistically better defined front of the signals in the BD counters due to more efficient light collection.
For the Be-Be counters, with the rise time as large as $15~ns$ and with significantly reduced statistics of collected photons, the time resolution is more dependent on the analysis method.

\section{Conclusion}

Scintillation counters coupled to SiPMs are excellent devices in trigger systems when a fast timing response is required. Nowadays, they are being used for testing several sub-detectors of the MPD experiment, and as trigger detectors in the BM@N experiment. In addition, they are considered as possible detectors in the Beam-Beam counter of the MPD.

The time resolution obtained in this study depends on the method chosen to determine the arrival time of the signals, for signals with shorter rise time this dependence is smaller.

As mentioned before, all these studies were performed with cosmic rays. An important complementary study and analysis of these detectors could be done using a particle beam in a high energy physics lab facilities; this will happen in the near future.

\acknowledgments{

The authors would like to thank the FFD group of the Laboratory of the High Energy Physics at the \textbf{J}oint \textbf{I}nstitute for \textbf{N}uclear \textbf{R}esearch~(JINR), and in particular to Sergey Sedykh, for his valuable help. Also, special thanks to Victor Rogov for helping with the experimental setup and SiPM electronics.

Marco thanks the JINR for the supporting received to join in the majestic Summer Student Program 2018, where part of this study was developed.

This work is supported by CONACYT~(Consejo Nacional de Ciencia y Tecnolog\'ia) under the project A1-S-23238.

}

\end{document}